\documentclass[12pt]{article}
\usepackage{epsfig}
\usepackage{amsmath}
\usepackage{hhline}
\usepackage{amssymb}
\usepackage{times}

\newlength{\dinwidth}
\newlength{\dinmargin}
\begin{document}  
\newcommand{\sleq}{\raisebox{-0.5mm}{$\stackrel{<}{\scriptstyle{\sim}}$}}
\newcommand{\sgeq}{\raisebox{-0.5mm}{$\stackrel{>}{\scriptstyle{\sim}}$}}
\newcommand{\Apom}{A_{\pom}}
\newcommand{\Areg}{A_{\reg}}
\newcommand{\alphapom}{\alpha_{_{I\!\!P}}}
\newcommand{\alphareg}{\alpha_{_{I\!\!R}}}

\newcommand{\pom}{{I\!\!P}}
\newcommand{\reg}{{I\!\!R}}
\newcommand{\slowpi}{\pi_{\mathit{slow}}}
\newcommand{\fiidiii}{F_2^{D(3)}}
\newcommand{\fiidiiiarg}{\fiidiii\,(\beta,\,Q^2,\,x)}
\newcommand{\n}{1.19\pm 0.06 (stat.) \pm0.07 (syst.)}
\newcommand{\nz}{1.30\pm 0.08 (stat.)^{+0.08}_{-0.14} (syst.)}
\newcommand{\fiidiiiful}{F_2^{D(4)}\,(\beta,\,Q^2,\,x,\,t)}
\newcommand{\fiipom}{\tilde F_2^D}
\newcommand{\ALPHA}{1.10\pm0.03 (stat.) \pm0.04 (syst.)}
\newcommand{\ALPHAZ}{1.15\pm0.04 (stat.)^{+0.04}_{-0.07} (syst.)}
\newcommand{\fiipomarg}{\fiipom\,(\beta,\,Q^2)}
\newcommand{\pomflux}{f_{\pom / p}}
\newcommand{\nxpom}{1.19\pm 0.06 (stat.) \pm0.07 (syst.)}
\newcommand {\gapprox}
   {\raisebox{-0.7ex}{$\stackrel {\textstyle>}{\sim}$}}
\newcommand {\lapprox}
   {\raisebox{-0.7ex}{$\stackrel {\textstyle<}{\sim}$}}
\def\gsim{\,\lower.25ex\hbox{$\scriptstyle\sim$}\kern-1.30ex%
\raise 0.55ex\hbox{$\scriptstyle >$}\,}
\def\lsim{\,\lower.25ex\hbox{$\scriptstyle\sim$}\kern-1.30ex%
\raise 0.55ex\hbox{$\scriptstyle <$}\,}
\newcommand{\pomfluxarg}{f_{\pom / p}\,(x_\pom)}
\newcommand{\dsf}{\mbox{$F_2^{D(3)}$}}
\newcommand{\dsfva}{\mbox{$F_2^{D(3)}(\beta,Q^2,x_{I\!\!P})$}}
\newcommand{\dsfvb}{\mbox{$F_2^{D(3)}(\beta,Q^2,x)$}}
\newcommand{\dsfpom}{$F_2^{I\!\!P}$}
\newcommand{\gap}{\stackrel{>}{\sim}}
\newcommand{\lap}{\stackrel{<}{\sim}}
\newcommand{\fem}{$F_2^{em}$}
\newcommand{\tsnmp}{$\tilde{\sigma}_{NC}(e^{\mp})$}
\newcommand{\tsnm}{$\tilde{\sigma}_{NC}(e^-)$}
\newcommand{\tsnp}{$\tilde{\sigma}_{NC}(e^+)$}
\newcommand{\st}{$\star$}
\newcommand{\sst}{$\star \star$}
\newcommand{\ssst}{$\star \star \star$}
\newcommand{\sssst}{$\star \star \star \star$}
\newcommand{\tw}{\theta_W}
\newcommand{\sw}{\sin{\theta_W}}
\newcommand{\cw}{\cos{\theta_W}}
\newcommand{\sww}{\sin^2{\theta_W}}
\newcommand{\cww}{\cos^2{\theta_W}}
\newcommand{\trm}{m_{\perp}}
\newcommand{\trp}{p_{\perp}}
\newcommand{\trmm}{m_{\perp}^2}
\newcommand{\trpp}{p_{\perp}^2}
\newcommand{\alp}{\alpha_s}

\newcommand{\alps}{\alpha_s}
\newcommand{\sqrts}{$\sqrt{s}$}
\newcommand{\LO}{$O(\alpha_s^0)$}
\newcommand{\Oa}{$O(\alpha_s)$}
\newcommand{\Oaa}{$O(\alpha_s^2)$}
\newcommand{\PT}{p_{\perp}}
\newcommand{\JPSI}{J/\psi}
\newcommand{\sh}{\hat{s}}
\newcommand{\uh}{\hat{u}}
\newcommand{\MP}{m_{J/\psi}}
\newcommand{\PO}{I\!\!P}
\newcommand{\xbj}{x}
\newcommand{\xpom}{x_{\PO}}
\newcommand{\ttbs}{\char'134}
\newcommand{\xpomlo}{3\times10^{-4}}  
\newcommand{\xpomup}{0.05}  
\newcommand{\dgr}{^\circ}
\newcommand{\pbarnt}{\,\mbox{{\rm pb$^{-1}$}}}
\newcommand{\gev}{\,\mbox{GeV}}
\newcommand{\WBoson}{\mbox{$W$}}
\newcommand{\fbarn}{\,\mbox{{\rm fb}}}
\newcommand{\fbarnt}{\,\mbox{{\rm fb$^{-1}$}}}
%
%
\newcommand{\qsq}{\ensuremath{Q^2} }
\newcommand{\gevsq}{\ensuremath{\mathrm{GeV}^2} }
\newcommand{\et}{\ensuremath{E_t^*} }
\newcommand{\rap}{\ensuremath{\eta^*} }
\newcommand{\gp}{\ensuremath{\gamma^*}p }
\newcommand{\dsiget}{\ensuremath{{\rm d}\sigma_{ep}/{\rm d}E_t^*} }
\newcommand{\dsigrap}{\ensuremath{{\rm d}\sigma_{ep}/{\rm d}\eta^*} }
\def\Journal#1#2#3#4{{#1} {\bf #2} (#3) #4}
\def\NCA{\em Nuovo Cimento}
\def\NIM{\em Nucl. Instrum. Methods}
\def\NIMA{{\em Nucl. Instrum. Methods} {\bf A}}
\def\NPB{{\em Nucl. Phys.}   {\bf B}}
\def\PLB{{\em Phys. Lett.}   {\bf B}}
\def\PRL{\em Phys. Rev. Lett.}
\def\PRD{{\em Phys. Rev.}    {\bf D}}
\def\ZPC{{\em Z. Phys.}      {\bf C}}
\def\EJC{{\em Eur. Phys. J.} {\bf C}}
\def\CPC{\em Comp. Phys. Commun.}
\def\PR{{\em Phys. Rep.}}
\def\MPL{{\em Mod. Phys. Lett.} A}
\def\AIP{{\em AIP Conf. Proc.}}

\def\st{\scriptstyle}
\def\sst{\scriptscriptstyle}
\def\mco{\multicolumn}
\def\epp{\epsilon^{\prime}}
\def\vep{\varepsilon}
\def\ra{\rightarrow}
\def\ppg{\pi^+\pi^-\gamma}
\def\vp{{\bf p}}
\def\ko{K^0}
\def\kb{\bar{K^0}}
\def\al{\alpha}
\def\ab{\bar{\alpha}}
\def\be{\begin{equation}}
\def\ee{\end{equation}}
\def\bea{\begin{eqnarray}}
\def\eea{\end{eqnarray}}
\def\CPbar{\hbox{{\rm CP}\hskip-1.80em{/}}}

\def\lapproxeq{\lower .7ex\hbox{$\;\stackrel{\textstyle <}{\sim}\;$}}
\def\gapproxeq{\lower .7ex\hbox{$\;\stackrel{\textstyle >}{\sim}\;$}}
\def\btop{\mathchar"1339}
\def\bmid{\mathchar"133D}
\def\bbot{\mathchar"133B}

\begin{titlepage}
\begin{flushleft}
DESY-98-209b \hfill ISSN 0418-9833\\
December 1998
\end{flushleft}

\vspace*{2cm} 

\begin{center}
\begin{LARGE}

{\bf\boldmath 
Diffraction and Low-$Q^2$ Physics \\
Including Two-Photon Physics}
\end{LARGE}

\vspace{2cm}

\begin{large}
Martin Erdmann \\
Universit\"at Karlsruhe, Engesserstr. 7, D-76128 Karlsruhe \\
E-mail: Martin.Erdmann@desy.de

\end{large}
\end{center}

\vspace{1cm}

\begin{abstract}
Recent experimental results on the partonic structure of the photon and on the 
color singlet exchange in strong interaction processes are reviewed. 
At the LEP $e^+e^-$  and HERA $ep$ colliders, complementary and consistent 
measurements have been achieved on the quark-gluon structure of quasi-real 
and virtual photons. 
At the HERA $ep$ and Tevatron $\bar{p}p$ colliders, the quark-gluon 
configuration of the diffractive exchange is consistently found to have a large 
gluon component. 
The rate of diffractive interactions observed by the HERA and 
Tevatron experiments, however, is largely different and challenges
explanation.
\end{abstract}
\vspace{4cm}

\noindent
{\em Invited plenary talk at the XXIX International Conference
on High Energy Physics, Vancouver, B.C. Canada (1998)}

\end{titlepage}

\section{The Partonic Structure of the Photon}\label{sec:photon}

\noindent
The motivation behind studying the structure of the photon results from the 
interest in understanding the formation of hadronic matter. 
Permitted by the Heisenberg uncertainty relation, the photon can fluctuate for 
some time into a quark--anti-quark state.  
This fluctuation can be disturbed, e.g., by an electron or proton probe 
which allows the density of quarks and gluons of the partonic state of the photon 
to be determined. 

\begin{figure}[hht]
\center\epsfig{file=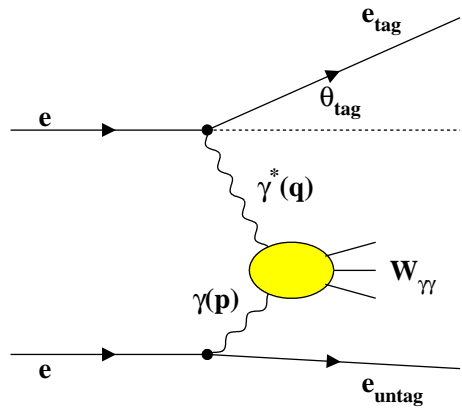,width=6.0cm}
\caption{Feynman diagram of deep inelastic electron--photon 
scattering:
the partonic structure of the quasi-real photon from the
untagged lepton is probed by the virtual photon from
the tagged electron.
}
\label{fig:diseg}
\end{figure}

\begin{figure}[hht]
\center\epsfig{file=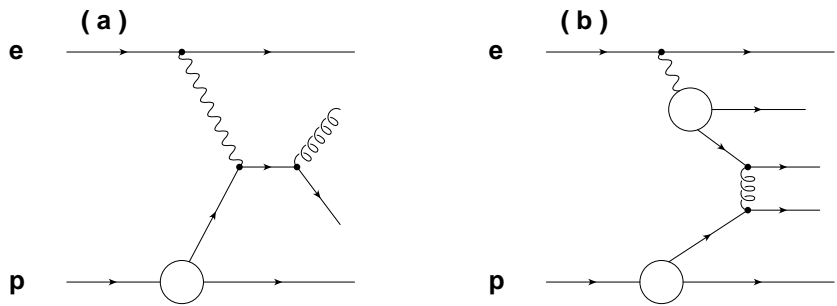,width=11.cm}
\caption{Examples of Feynman diagrams for photoproduction of jets in $ep$
collisions in leading order QCD:
a) direct photon--proton process, b) resolved photon--proton process.
}
\label{fig:jetgp}
\end{figure}

At the LEP $e^+e^-$ and HERA $ep$ colliders, photons are emitted by the 
leptons which gives access to the partonic structure of almost real photons 
\cite{erdmann} as well as highly virtual photons. 
The measurements to obtain information on the partonic state of the 
photons discussed here are
\begin{enumerate}
\item the photon structure function from deep inelastic electron--photon 
scattering (Fig.~\ref{fig:diseg}),
\item jet and particle cross sections (e.g. Fig.~\ref{fig:jetgp}), and
\item the total photon--photon cross section. 
\end{enumerate}

\subsection{Measurements Related to the Quark Distributions of Quasi-Real Photons}
\label{subsec:g-quark}

\begin{figure}[hht]
\center
\epsfig{file=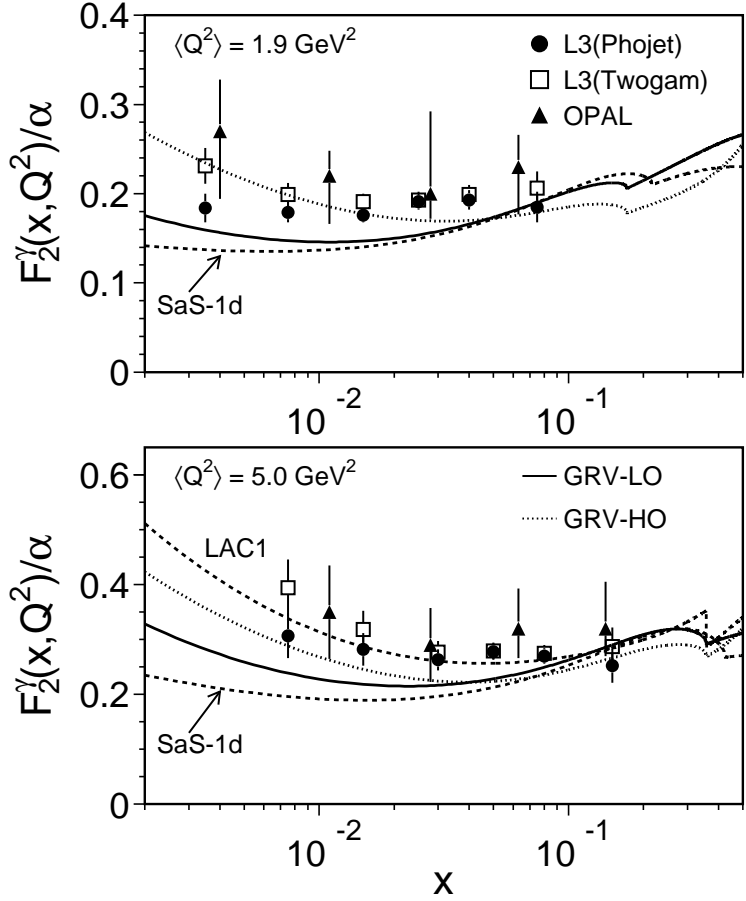,width=12cm}
\caption{
The photon structure function $F_2^\gamma$, measured in 
two--photon collision at LEP, is shown as a 
function of the parton fractional momentum $x$ in two bins of
the virtuality $Q^2$ of the probing photon.
The squared symbols and the circles represent the measurements
of the L3 experiment using two different Monte Carlo generators
for correcting detector effects.
For comparison, previous results of the OPAL experiment are shown
(triangle symbols).
The curves represent different parameterizations of the parton
distributions in the photon.}
\label{fig:f2-x}
\end{figure}

\noindent
New $F_2^\gamma$ structure function measurements 
have been performed in the interesting 
region of small parton momenta $x \sim 10^{-2}$ by the L3 collaboration
~\cite{l3-f2}. 
$F_2^\gamma$ is determined from the measurement of the double differential 
inclusive cross section 
\begin{equation}
\frac{d^2\sigma}{dx \, dQ^2} =
\frac{2 \pi \alpha^2}{x \, Q^4} \left( 1 + ( 1 - y )^2 \right) \;
F_2^\gamma(x, Q^2) \; ,
\label{eq:f2gamma}
\end{equation}
where $\alpha$ is the electro-magnetic coupling constant, 
$Q^2$ denotes the virtuality of the probing photon and gives the 
resolution scale of the process, and $y$ is the inelasticity 
$y=Q^2/(x s_{e\gamma})$.
In Fig.~\ref{fig:f2-x},
the $x$ dependence of $F_2^\gamma$ is shown in two bins of $Q^2$.

A major challenge in this analysis is the determination of $x$:
since the lepton that emitted the target photon remains undetected, the energy of 
the target has to be determined from the hadronic final state. 
Using a new improved reconstruction method for $x$,
two results for $F_2^\gamma$ are presented by the L3 collaboration using two 
different Monte Carlo generators for the correction of detector effects
(Phojet ~\cite{phojet}, Twogam ~\cite{twogam}). 
These two data sets demonstrate that over a large region in $x$ the structure 
function result does not depend on the details of simulating the hadronic 
final state. 
Only below $x \sim 10^{-2}$ this limitation becomes sizable. 

In the same figure, previous results of the OPAL collaboration are shown
~\cite{opal-f2}.
Within the errors, good agreement is observed between the two experiments.
Also shown are different parameterizations of the quark density in 
the photon demonstrating that the data give new information on the quark 
distributions at low $x$ (LAC ~\cite{lac}, GRV ~\cite{ggrv}, SaS ~\cite{sas}). 
Scaling violations caused by gluon emission off the quark before the 
scattering process occurs results in a rise of $F_2^\gamma$ 
below a small value of $x$.
The data are not yet precise enough to confirm or reject such a rise 
at $x \sim 10^{-2}$.

\begin{figure}[hht]
\center\epsfig{file=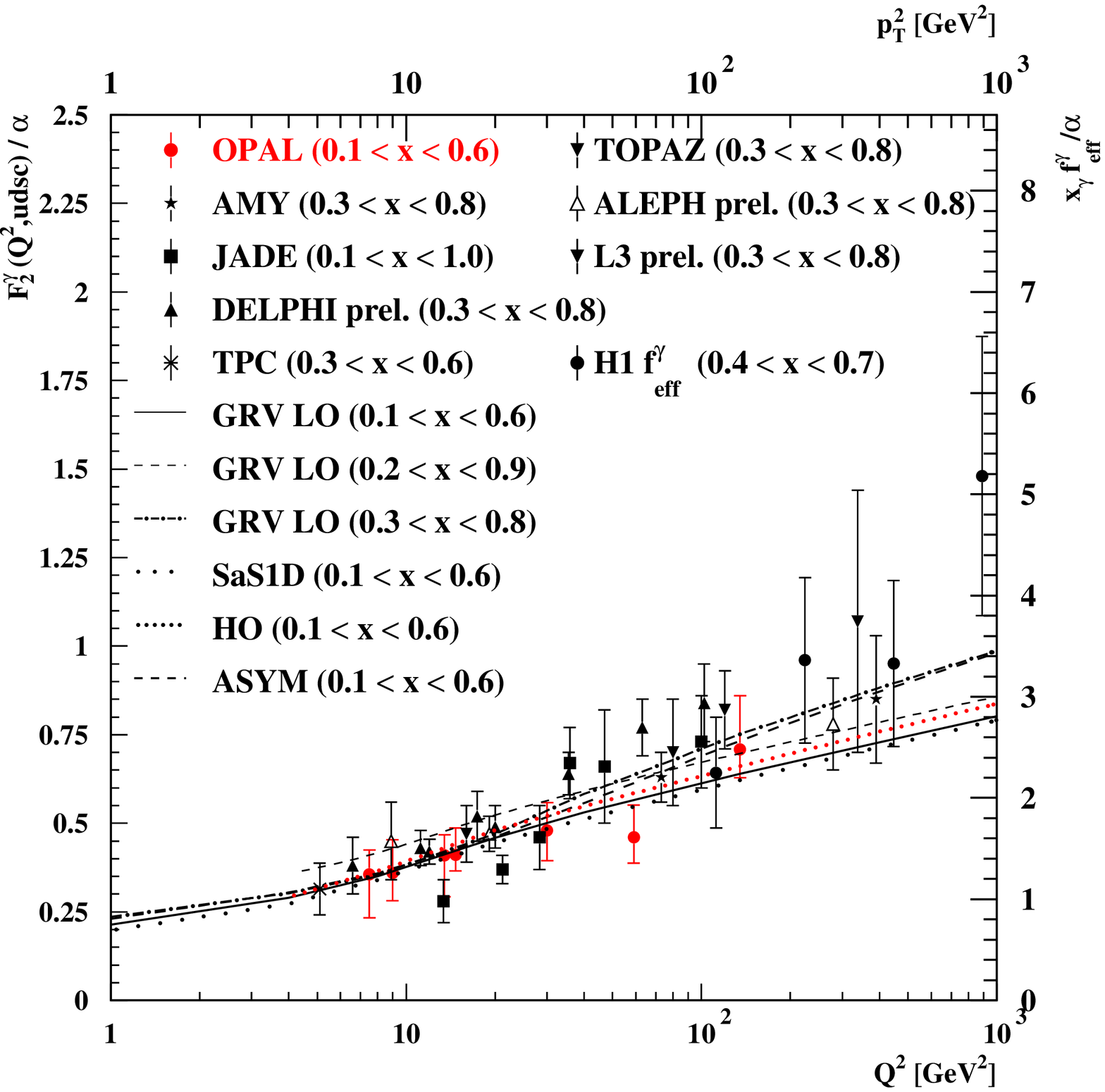,width=11.0cm}
\caption{The structure function $F_2$ of the photon is 
shown as a function of the
virtuality $Q^2$ of the probing photon for parton fractional
momenta around $x\sim 0.5$.
Measurements of the photon structure function $F_2^\gamma$ 
from $e^+e^-$ data are shown in comparison with an effective
parton distribution extracted from photoproduction of di-jets
in $ep$ collisions (H1 data).
The curves represent different parameterizations of the parton
distributions of the photon.}
\label{fig:f2-q2-g}
\end{figure}

\begin{figure}[hht]
\center\epsfig{file=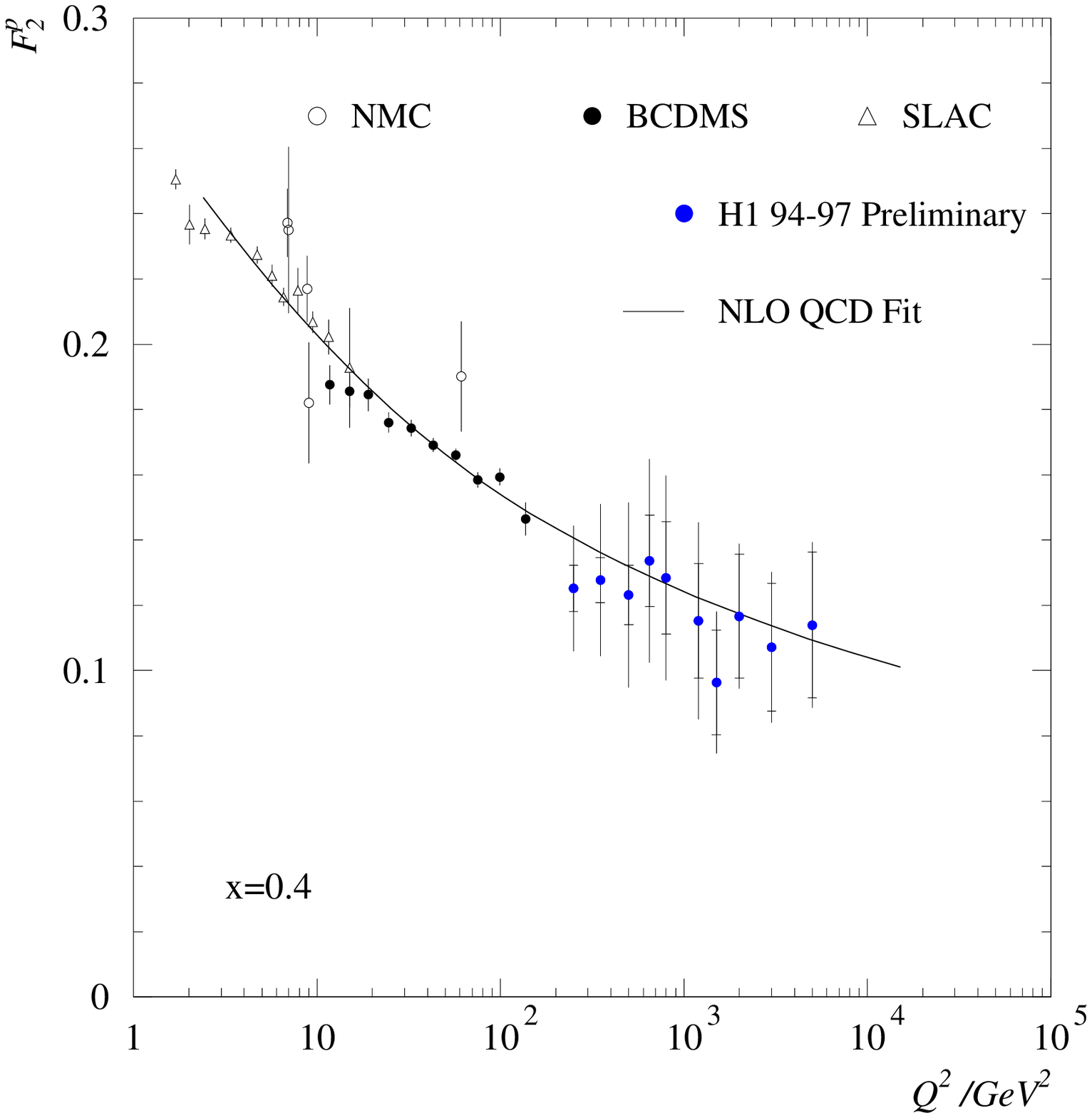,width=10cm}
\caption{The structure function $F_2$ of the proton is shown 
as a function of the
virtuality $Q^2$ of the probing photon for the parton fractional
momentum $x=0.4$ from fixed target data and preliminary H1 data.
}
\label{fig:f2-q2-p}
\end{figure}

In the momentum region around $x \sim 0.5$, where the quark and the 
anti-quark each carry half of the photon energy, results on the structure function 
$F_2^\gamma$ exist from many experiments. 
A compilation of these measurements is shown in 
Fig.~\ref{fig:f2-q2-g} as a function of the resolution scale 
$Q^2$ ~\cite{nisius}. 
The data are compatible with an increasing quark density in the photon as 
$Q^2$ increases. 
This $Q^2$ dependence is very different from that of hadronic 
structure functions at large $x$ and is expected by perturbative QCD
(Fig.~\ref{fig:f2-q2-p} and discussion in Section~\ref{subsec:cse}):
the splitting of the photon into a quark-anti-quark pair gives rise to the 
probability $f_{q/\gamma}$ of finding a quark in the photon to increase as 
\begin{equation}
f_{q/\gamma}\sim \ln{\frac{Q^2}{\Lambda_{QCD}^2}} 
\end{equation}
in leading order.

In the same figure an effective parton distribution $x\tilde{f}_\gamma$ of the 
photon is shown which has been extracted from di-jet measurements in 
photon--proton collisions by the H1 collaboration ~\cite{h1-effpdf}. 
This effective parton distribution combines the quark and the gluon densities of 
the photon with a weight of color factors ~\cite{combridge}: 
\begin{equation}
x \tilde{f}_\gamma = x \; (f_{q/\gamma} + \frac{9}{4} \, f_{g/\gamma}) \; .
\label{eq:effpdf}
\end{equation}
The vertical scale for $x \tilde{f}_\gamma$ 
on the right side of Fig.\ref{fig:f2-q2-g}
has been adjusted relative to the $F_2^ \gamma$ scale, since 
in contrast to the $F_2^\gamma$ measurements the jet processes are 
independent of the electric charges of the quarks. 
The relevant resolution scale is the transverse momentum $p_t^2$ of the scattered 
partons which is here taken to have the same resolution power as $Q^2$. 
The results of the di-jet measurements are in good agreement with the $F_2^\gamma$ data. 
The jet data probe the partons of the photon at large resolution scales and
compete well in precision with the $F_2^\gamma$ measurements. 

The quark density close to the kinematic limit $x \sim 1$ is analysed in 
photoproduction of two jets. 
Here the contributions of the direct and resolved photon--proton 
processes need to be understood (Fig.~\ref{fig:jetgp}).
They differ in their matrix elements and therefore in the distribution of the 
parton scattering angle $\theta^*$. 

\begin{figure}[hht]
\setlength{\unitlength}{1cm}
\begin{picture}(5.0,10.0)
\put(0.5,-0.9)
{\epsfig{file=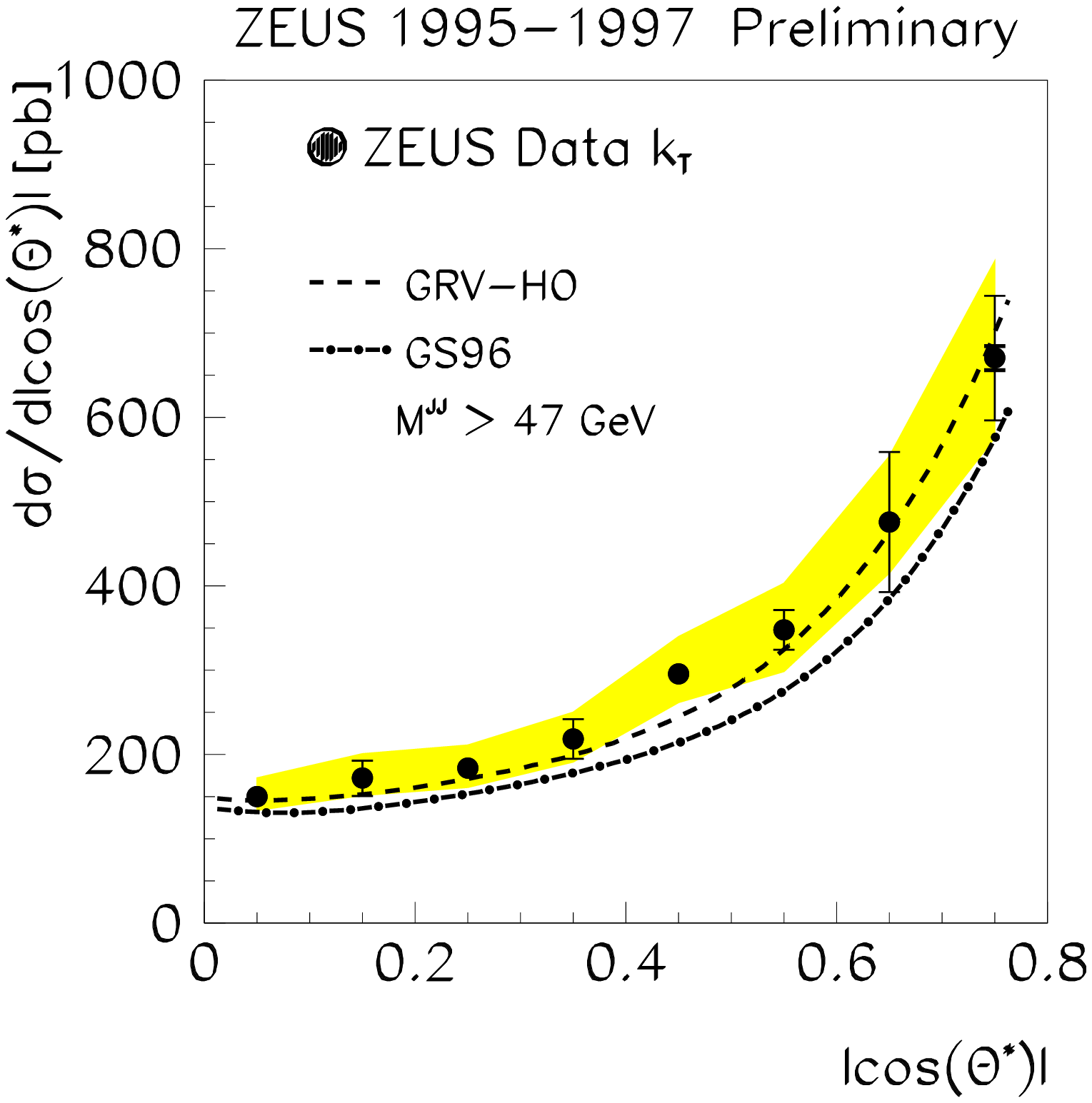,width=12.5cm}}
\end{picture}
\caption{The di-jet cross section in $ep$ collisions involving quasi-real photons
is shown differentially in terms of the cosine of the parton
scattering angle at large di-jet mass above $47$ GeV from ZEUS data.
The curves represent next-to-leading QCD calculations using different parton
distributions of the photon.
}
\label{fig:costh}
\end{figure}

In Fig.~\ref{fig:costh},
a new di-jet cross section measurement of the ZEUS collaboration is 
shown differentially in $\vert \cos{\theta^*} \vert$ for large di-jet masses and 
correspondingly large $x$ ~\cite{zeus-costh}.  
Also shown are next-to-leading order QCD calculations ~\cite{klasen}
using two different 
parton parameterizations of the photon (GRV ~\cite{ggrv}, GS ~\cite{gs96}). 
The direct photon contribution 
(not shown in the figure)
is not sufficient to describe the measured jet 
cross section either in shape or in the absolute normalization.  
Contributions of resolved photon processes are required to describe the data 
which are sufficiently precise to discriminate different parton parameterizations 
of the photon at large $x$.

\subsection{Measurements Related to the Gluon Distribution of Quasi-Real Photons}
\label{subsec:g-gluon}

\begin{figure}[hht]
\center
\epsfig{file=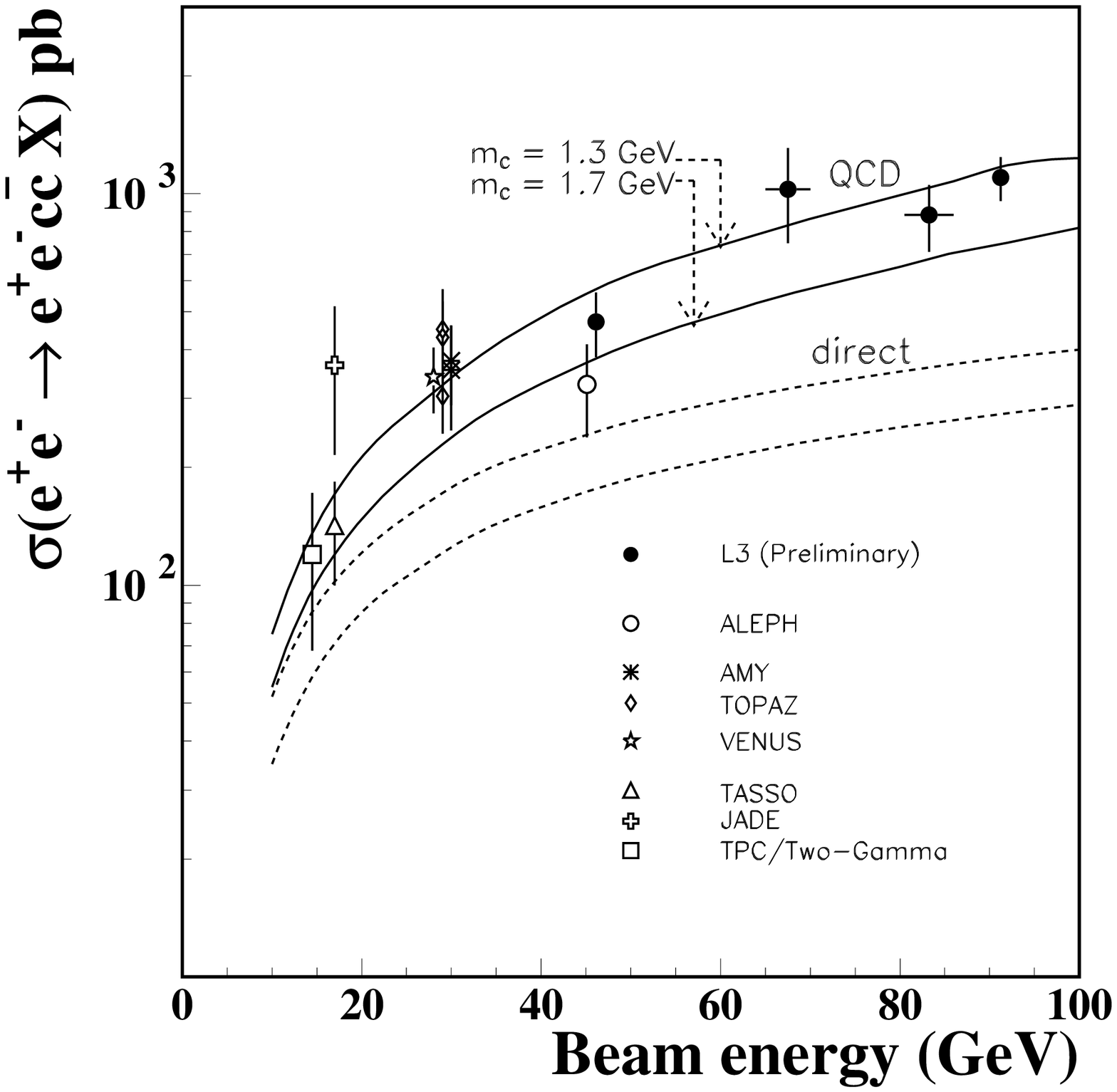,width=10cm}
\caption{Measurements of the total charm production cross sections from
two--photon collisions are shown as a function of the lepton beam
energy (L3 experiment).
The full curves represent next-to-leading QCD calculations using the GRV parton
distribution functions of the photon and different values for 
the charm mass.
The direct photon contribution is shown separately (dashed curves).
}
\label{fig:charm}
\end{figure}

\noindent
New measurements of the inclusive charm production cross section at the large 
LEP beam energies are shown in Fig.~\ref{fig:charm} by the L3 collaboration
~\cite{l3-charm}.  
The cross section has been determined using semi-leptonic charm decays in the
electron and muon channels.
In the same figure, next-to-leading order QCD calculations ~\cite{nlo-charm}
using two different 
charm masses and the GRV parameterization ~\cite{ggrv}
of the parton distributions in the photons are shown. 
The dominant contribution to the cross section results from gluon induced 
processes with an average gluon momentum as small as 
$\langle x \rangle \sim 0.03$ ~\cite{andreev}. 

\begin{figure}[hht]
\center{\epsfig{file=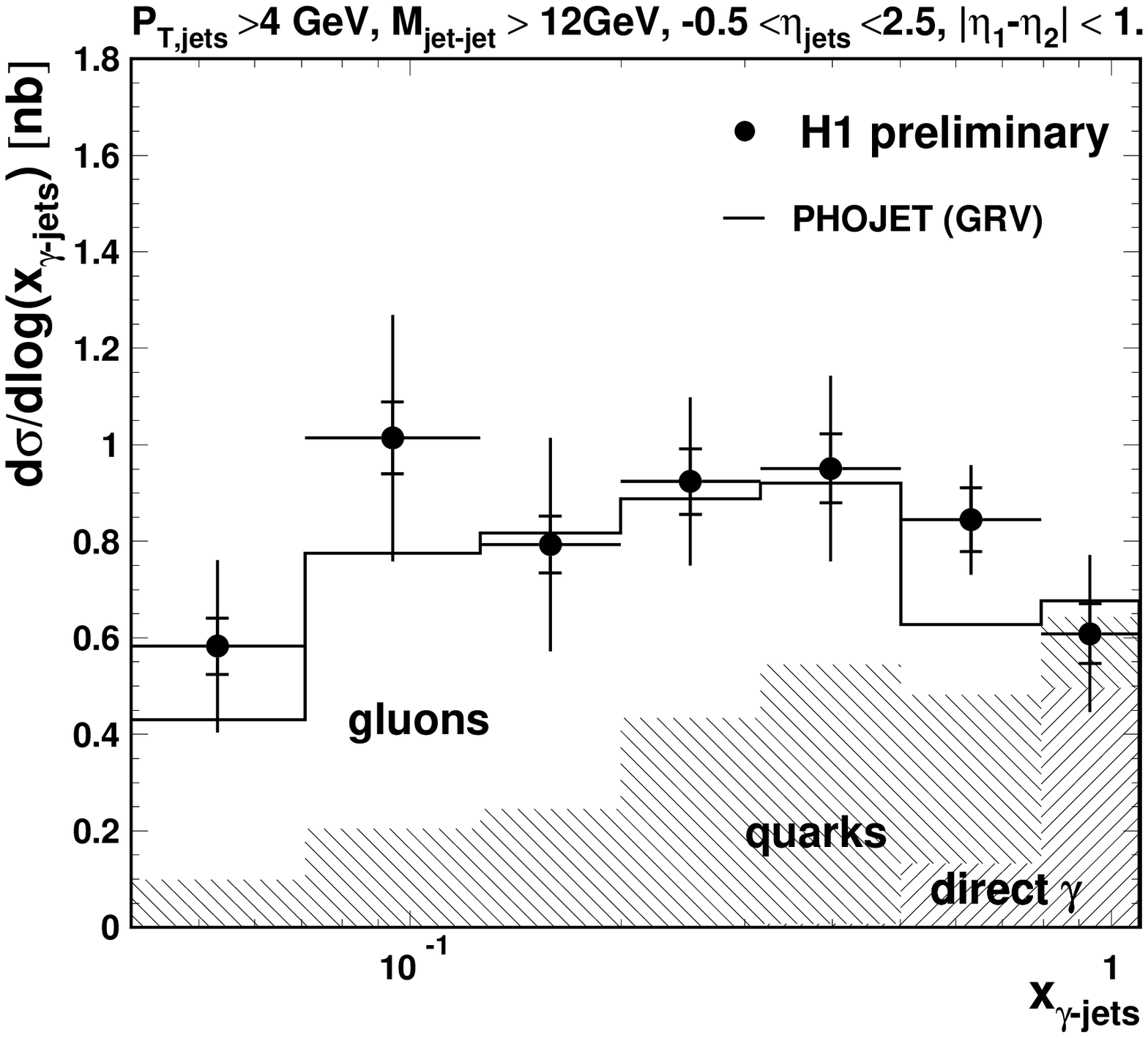,width=10cm}}
\caption{Photoproduction of di-jets in $ep$ collisions is shown as 
a function of the parton fractional momentum $x$ from H1 data.
The cross section measurement is compared to a leading-order QCD
calculation showing above the quark and direct photon contributions the
gluon component of the photon at small $x$ (the histograms are calculated
using the GRV parameterizations of the partons in the photon).}
\label{fig:xgamma}
\end{figure}

Also di-jet data are used to access the low-$x$ gluon distributions of the photon. 
In Fig.~\ref{fig:xgamma},
a new measurement of the di-jet cross section is shown as a function 
of the parton momentum $x$ by the H1 collaboration ~\cite{h1-xgamma}. 
The histograms represent a leading-order QCD calculation ~\cite{phojet}
showing the 
contributions of the direct photon-proton interactions and quark and gluon 
induced processes using 
the GRV parton parameterizations for the photon and the proton. 

Both the charm and di-jet measurements give compatible conclusions on the 
low-$x $ gluon density of the photon and are precise to the level of $30\%$. 

\begin{figure}[hht]
\center
\epsfig{file=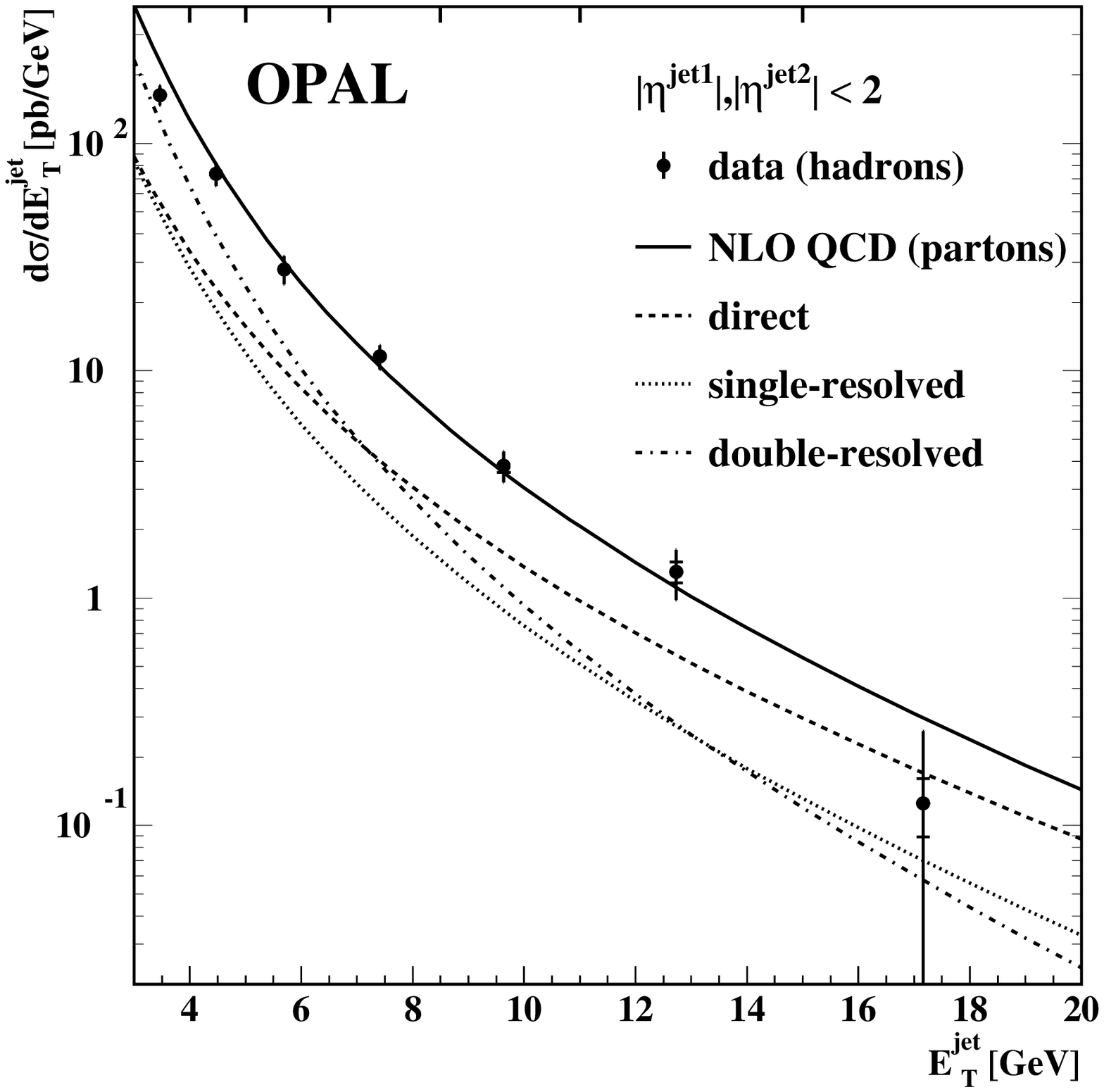,width=10cm}
\caption{The di-jet cross sections from two--photon processes in $e^+e^-$
collisions is shown as a function of the transverse jet energy from
OPAL data.
The curves represent next-to-leading QCD calculations of the different photon
contributions using the GRV parton distribution functions of the photon.
The labels refer to direct photon--photon interactions via quark exchange
(direct), processes where one photon interacts directly with a parton
of the other photon (single resolved) and processes which involve partons 
of both photons (double resolved).
}
\label{fig:opal-pt}
\end{figure}

New information on the gluon distribution of the photon results 
from di-jet production in photon-photon collisions which has been measured by the 
OPAL collaboration ~\cite{opal-dijet}. 
In Fig.~\ref{fig:opal-pt}, the cross section is  
shown differentially in the transverse jet energy $E_t^{jet}$.
At sufficiently large $E_t^{jet}$
the measurement can well be described by a next-to-leading order QCD
calculation
~\cite{kramer} using the parton distribution function of GRV ~\cite{ggrv}.

\begin{figure}[hht]
\setlength{\unitlength}{1cm}
\begin{picture}(5.0,10.0)
\put(0.0,0.0)
{\epsfig{file=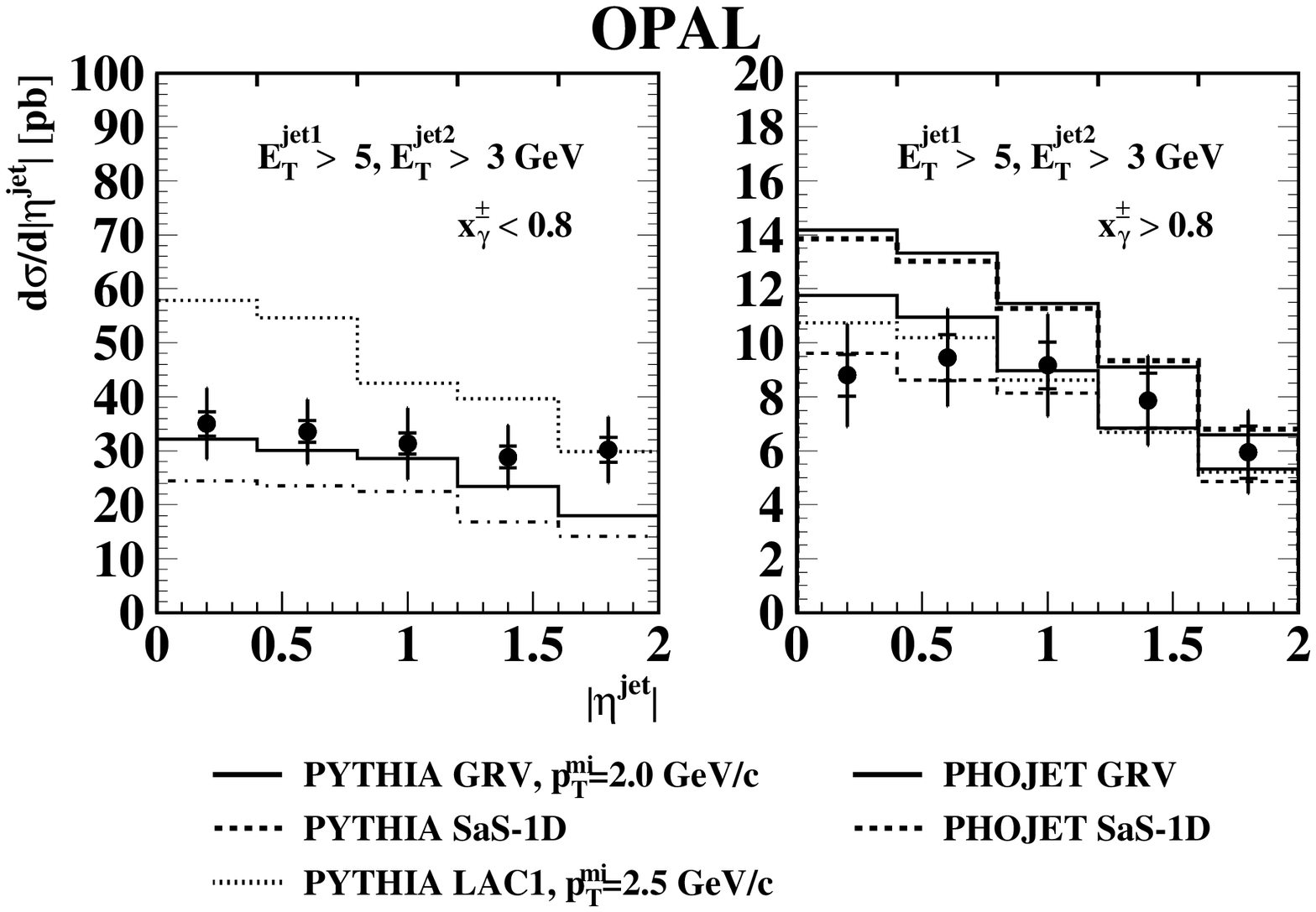,width=14cm}}
\end{picture}
\caption{ The di-jet cross sections from two--photon processes in $e^+e^-$
collisions is shown differentially in terms of the jet pseudo-rapidity
in two bins of the reconstructed parton fractional momentum $x$ of the photon
(OPAL experiment).
The histograms represent leading order QCD calculations of two Monte Carlo
generators using different parton distribution functions for the photon.}
\label{fig:opal-eta}
\end{figure}

In Fig.~\ref{fig:opal-eta},
the di-jet cross sections are shown differentially in the jet rapidity 
$\vert\eta^{jet}\vert$.
The data explore different regions of the parton fractional momentum 
$x>0.8, x<0.8$ with a precision of $\sim 20\%$.
They are compared to leading order QCD calculations (Phojet ~\cite{phojet},
Pythia ~\cite{pythia}) and 
discriminate different parameterizations of the gluon distributions of the 
photon (LAC ~\cite{lac}, GRV ~\cite{ggrv}, SaS ~\cite{sas}).

\newpage
\subsection{Parton Distributions of Virtual Photons}\label{subsec:g-virtual}

\noindent
The fluctuation of a virtual photon into a quark-anti-quark pair is suppressed by 
the photon virtuality $Q^2$. 
In comparison with real photons one therefore expects a smaller
probability of finding the virtual photon in a partonic state.
Also, there is less time to develop from the $q\bar{q}$ pair a vector meson 
bound state such
that the hadronic contributions to the virtual photon structure should be small. 

\begin{figure}[hht]
\setlength{\unitlength}{1cm}
\begin{picture}(5.0,9.0)
\put(1.,2.)
{a)}
\put(5.,2.)
{b)}
\put(0.,0.)
{\epsfig{file=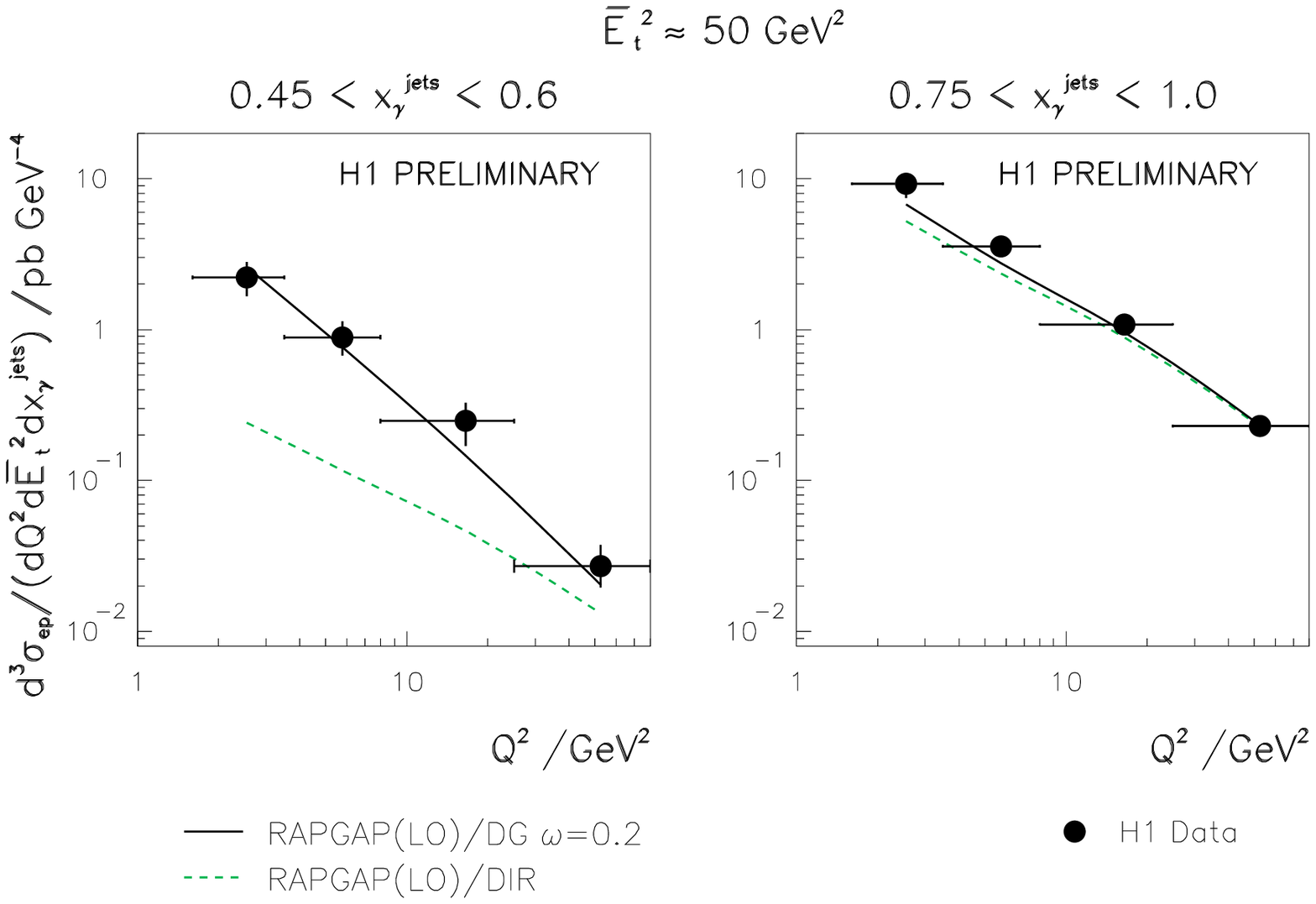,width=14cm}}
\end{picture}
\caption{The triple differential di-jet cross section from $ep$
collisions is shown as a function of the photon virtuality $Q^2$
for two bins of the fractional momentum $x$ of the parton from the photon
at fixed transverse jet energy $E_t^{jet}$ (H1 experiment).
The full curve is a leading order QCD calculation including the direct
photon--proton interactions (dashed curve) and resolved photon processes,
the latter reflecting the partonic structure of the virtual photon.}
\label{fig:virtual}
\end{figure}

In Fig.~\ref{fig:virtual},
the first triple-differential di-jet cross section is shown 
as a function of the 
photon virtuality $Q^2$ in two bins of the parton momentum $x$ for 
a fixed resolution scale $(E_t^{jet})^2 = 50$ GeV$^2$ ~\cite{h1-virtual}. 
The cross section measurement at $x \sim 1$ (Fig.~\ref{fig:virtual}b) 
is well described by a 
leading order QCD calculation using the direct photon-proton interaction 
processes only (dashed curve ~\cite{rapgap}). 
At $x \sim 0.5$ (Fig.~\ref{fig:virtual}a) 
the absolute cross section is found to be smaller 
compared to the measurement at $x \sim 1$ as expected from the short
fluctuation time of the photon. 
Here the direct photon contributions are not sufficient to describe the data at 
small $Q^2\sim 2$ GeV$^2$:
the di-jet process is able to resolve the partonic structure of the virtual photon. 
As $Q^2$ approaches the squared transverse energy of the jets of $(E_t^{jet})^2=50$ GeV$^2$, 
the resolution power of the di-jet process becomes insufficient for detecting the
fluctuations of the virtual photons.

\begin{figure}[hht]
\setlength{\unitlength}{1cm}
\begin{picture}(5.0,10.0)
\put(12.2,8.4)
{b)}
\put(4.5,8.4)
{a)}
\put(0.,0.)
{\epsfig{file=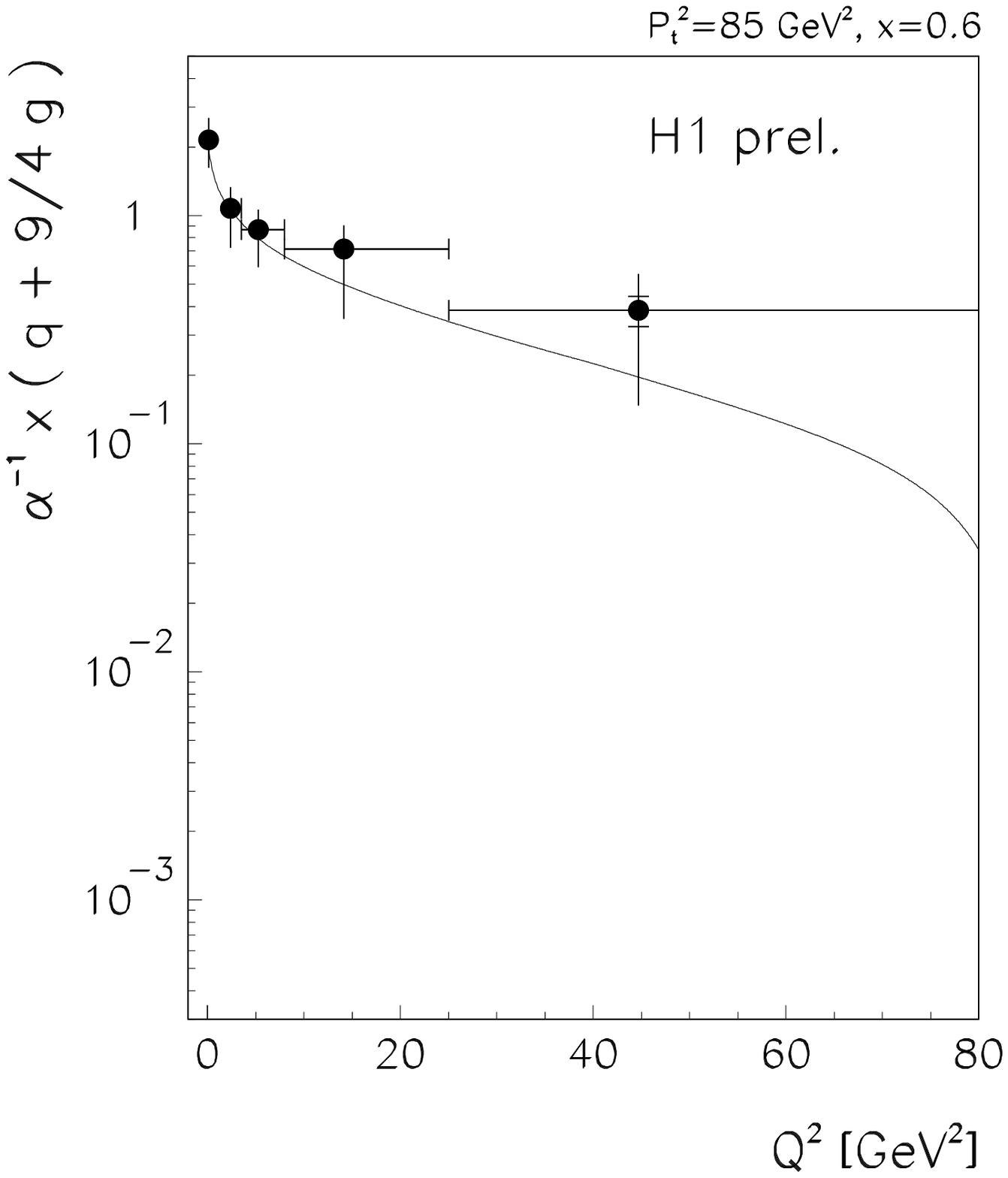,width=8.5cm}}
\put(8.,0.)
{\epsfig{file=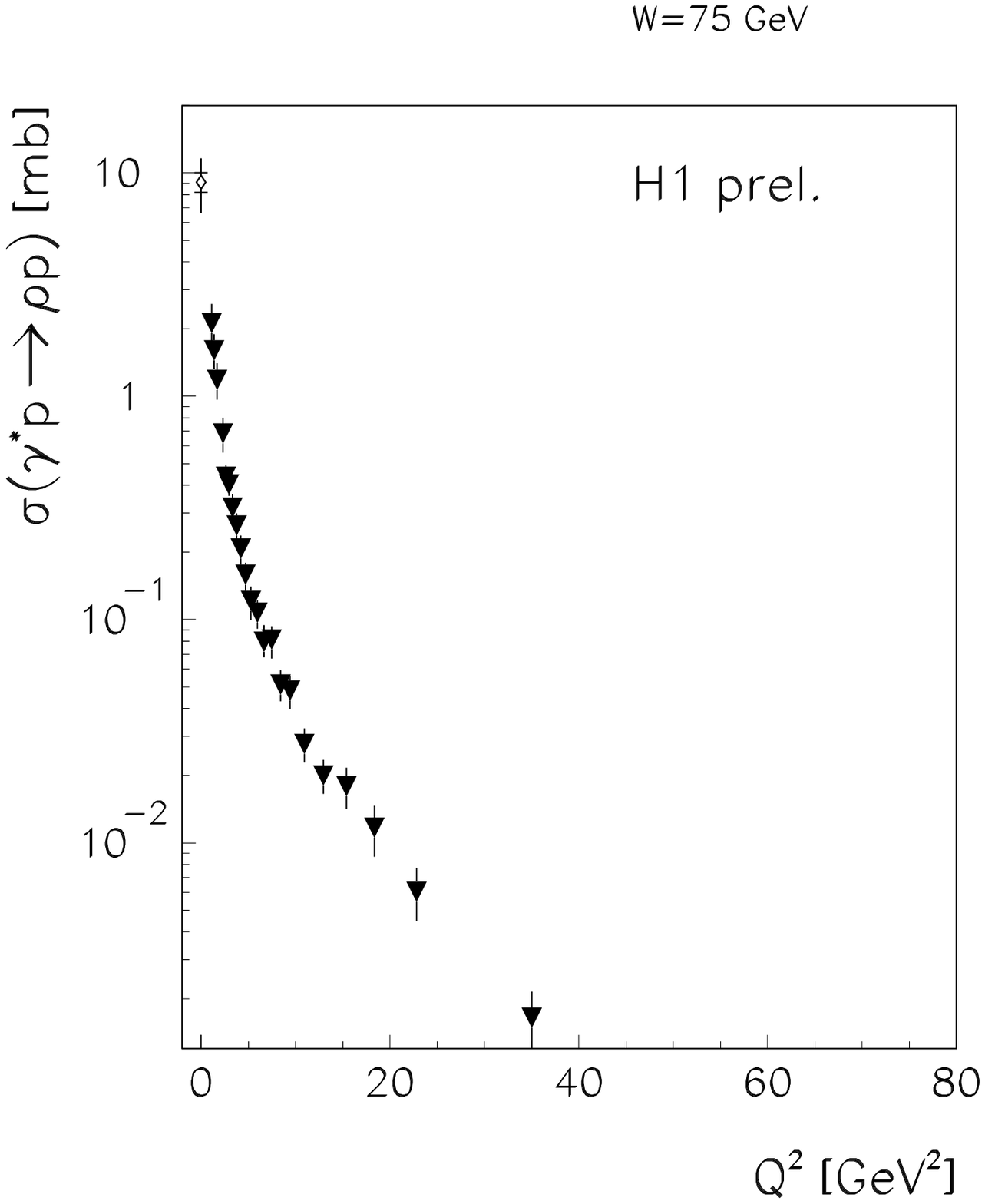,width=8.5cm}}
\end{picture}
\caption{
a) The effective parton distribution of virtual photons is shown as 
a function of the photon virtuality $Q^2$ for fixed parton
fractional momentum $x=0.6$ and scattered parton squared 
transverse momentum $p_t^2=85$ GeV$^2$ (H1 experiment).
The curve represents the SaS1d parameterization of the 
photon effective parton distribution.
b) The $\rho$ meson cross section is shown as a function of the photon
virtuality $Q^2$ for the photon-proton center of mass energy $W=75$ GeV.
}
\label{fig:effpdf}
\end{figure}

In analogy to the real photon case, eq.~(\ref{eq:effpdf}), 
an effective parton distribution for virtual photons 
$x \tilde{f}_{\gamma^*} = x (f_{q/\gamma^*} + 9/4 f_{g/\gamma^*})$ 
has been extracted from the data and 
is shown in Fig.~\ref{fig:effpdf}a
in the interval $0 \le Q^2\le 80$ GeV$^2$ for $x = 0.6$ and
$(E_t^{jet})^2=85$ GeV$^2$.
The partonic structure of the virtual photon is only slowly suppressed with the 
photon virtuality $Q^2$. 
Such a dependence is predicted by perturbative QCD:
in the region of $\Lambda_{QCD}^2 < Q^2 < (E_t^{jet})^2$ the 
probability of finding a quark in the virtual photon decreases logarithmically as 
$Q^2$ approaches the jet resolution scale: 
\begin{equation}
f_{q/\gamma^*} \sim \ln{\frac{(E_t^{jet})^2}{Q^2}} \; .
\end{equation}

The formation of a hadronic bound state from the $q\bar{q}$ pair 
of the photon can be studied with the production of $\rho$ mesons.
In Fig.~\ref{fig:effpdf}b, the $Q^2$ dependence of the $\rho$ 
cross section is shown which exhibits a fast decrease proportional 
to $(Q^2 + M_\rho^2)^{-n}$ with $n = 2.24\pm 0.09$ ~\cite{h1-rho}. 
As expected from the short 
photon fluctuation time into a quark-anti-quark pair, the 
probability to develop a hadronic bound state from the 
quark-anti-quark pair is highly suppressed. 
At sufficiently large $Q^2$, the partonic structure of the virtual photon can 
therefore be predicted by perturbative QCD. 
In Fig.~\ref{fig:effpdf}a, the full curve represents a QCD inspired model of 
the effective parton distribution of the virtual photon (SaS1d ~\cite{sasvirt})
which is in agreement with the measurement within the experimental errors.                  

\subsection{Total Photon-Photon Cross Section}\label{subsec:g-total}

\noindent
The total photon-photon cross section $\sigma_{\gamma\gamma}$
is dominated by soft 
scattering processes in which the photons develop a hadronic structure before 
the interaction occurs. 
A major challenge of this measurement is the understanding of the different 
contributions, the elastic, diffractive and non-diffractive processes. 
The visibility of the first two contributions in the detectors is small and requires 
reliable Monte Carlo generator calculations. 

\begin{figure}[hht]
\center
\epsfig{file=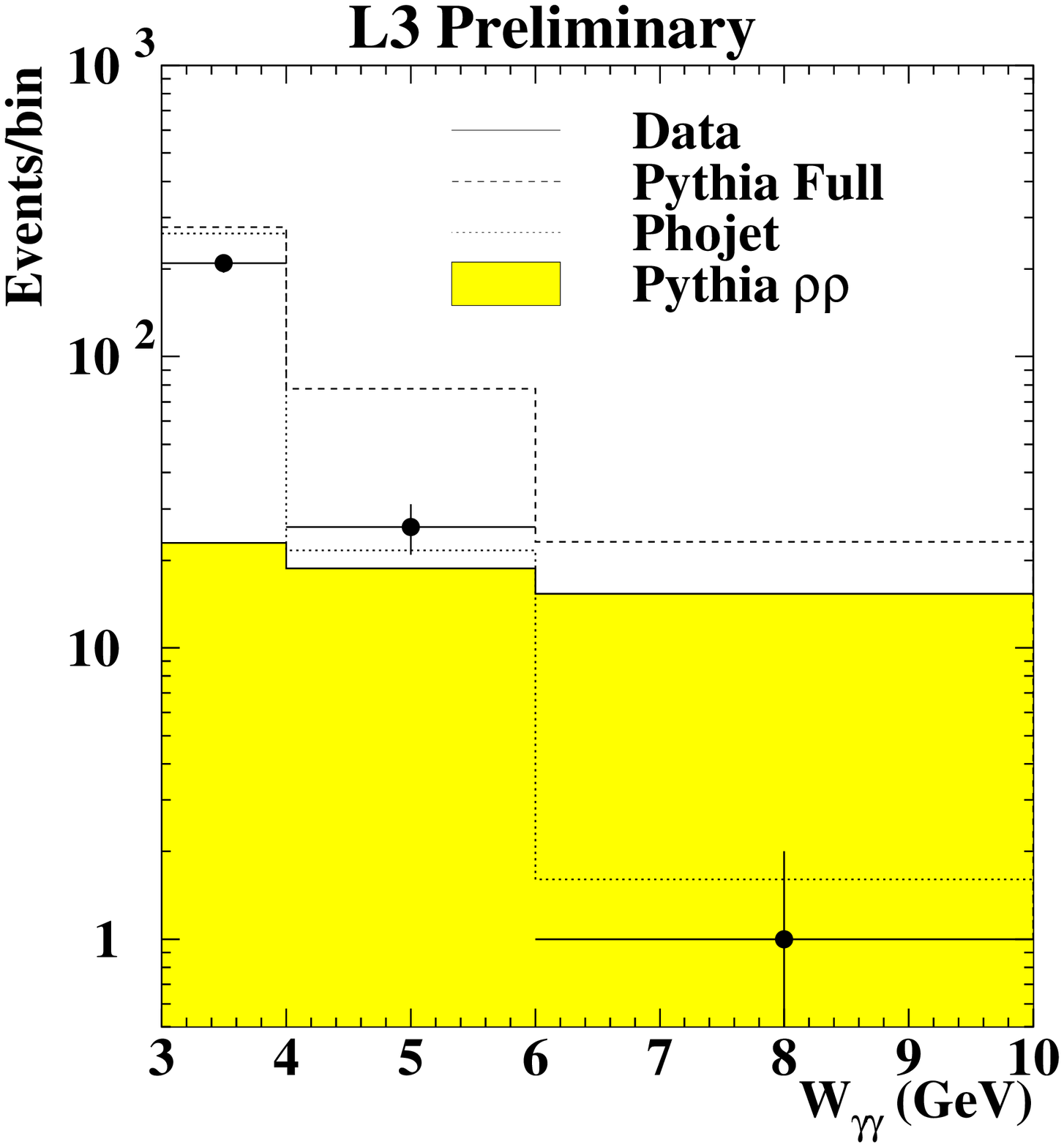,width=8.0cm}
\caption{
The number of events with exclusive 
four pion production from two-photon collisions in $e^+e^-$ scattering 
is shown as a function of the photon--photon center of mass energy 
$W_{\gamma\gamma}\equiv \sqrt{s_{\gamma\gamma}}$ (L3 experiment).
The data are compared with calculations of different Monte Carlo generators,
showing the total four-pion production (histograms) and 
the exclusive production two $\rho$ mesons (shaded histogram) separately.}
\label{fig:sigma-rr}
\end{figure}

Progress has recently been made by the L3 experiment which succeeded in
collecting a few hundred events of exclusive four pion production which
contains contributions of 
elastic double-$\rho$ production at center of 
mass energies below $10$ GeV (Fig.~\ref{fig:sigma-rr}) ~\cite{l3-sigma-gg}. 
These data test the two generator calculations shown 
(Phojet ~\cite{phojet}, Pythia ~\cite{pythia}). 

\begin{figure}[hht]
\setlength{\unitlength}{1cm}
\begin{picture}(10.0,8.4)
\put(4.,0.)
{\epsfig{file=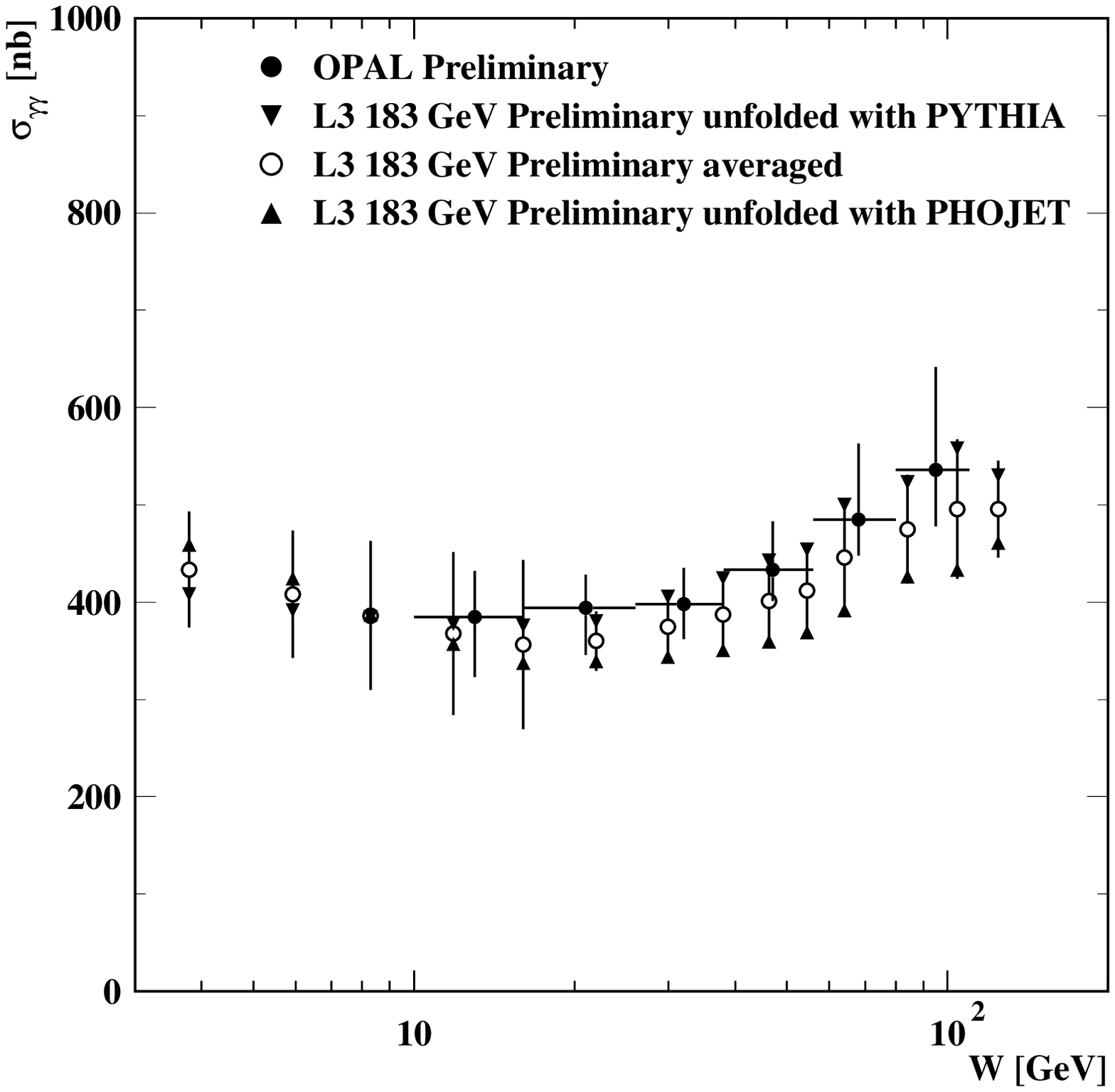,width=8.0cm}}
\end{picture}
\caption{The total photon-photon cross section is shown as a function of the photon--photon center of mass energy 
$W_{\gamma\gamma}\equiv \sqrt{s_{\gamma\gamma}}$. 
The preliminary measurements of the L3 
and OPAL 
experiments are compared 
using two different Monte Carlo generators for the detector corrections. }
\label{fig:sigma-gg}
\end{figure}

A new measurement of the total photon-photon cross 
section is shown in Fig.~\ref{fig:sigma-gg} using the two
different Monte Carlo generators (L3 collaboration ~\cite{l3-sigma-gg,ssr}). 
The data show a rise above $W\equiv\sqrt{s_{\gamma\gamma}} = 10$~GeV 
and are compatible within errors with the preliminary 
measurement of the OPAL collaboration ~\cite{opal-sigma-gg}.
This observed rise can be described by a power law 
$s_{\gamma\gamma}^\epsilon$ with 
\mbox{$\epsilon=0.158\pm0.006\pm0.028$ ~\cite{l3-sigma-gg}.}
The rise has the tendency to be stronger 
than expected from soft Pomeron exchange which successfully describes all 
hadron--hadron and photon--proton total cross sections with
\mbox{$\epsilon=0.095\pm 0.002$ ~\cite{epsilon}.}

\subsection{Summary 1: Photon}\label{subsec:g-summary}

\noindent
Improved knowledge on the partonic structure of real photons results from
\begin{itemize}
\item new 
structure function $F_ 2^ \gamma $ measurements at low parton fractional
momenta $x\sim 10^{-2}$, 
\item
di-jet cross section measurements at $x$ values down to $\sim 10^{-2}$ and 
high $x\rightarrow 1$ in photon-proton and photon-photon interactions, and 
\item
charm 
production in photon--photon processes at low $x\sim 10^{-2}$. 
\end{itemize}

For the first time the partonic structure of highly virtual photons 
$Q^2>1$ GeV$^2$ has been investigated in $ep$ collisions. 
The fluctuations of the virtual photon into a quark-anti-quark pair is only slowly 
suppressed with $Q^2$ and is compatible with a logarithmic decrease as 
predicted by perturbative QCD. 

The understanding of the total photon-photon cross section has improved by 
the detection of elastic $\rho$ production. 

Overall, the results on the photon 
obtained in $e^+e^-$ and $ep$ collisions complement
each other and are well compatible.
The precision of the measurements remains a challenge for the next few 
years in order to be well prepared for the linear collider.              

\section{Colour Singlet Exchange}

\noindent
A sizable fraction of strong interaction processes 
includes the exchange of colour singlet objects.
At the HERA $ep$ and Tevatron $\bar{p}p$ colliders, these objects
are emitted by the hadrons and may involve the exchange of quantum 
numbers (meson exchange) or may not (diffractive processes).

A handle on the type of the interaction process is given, e.g., by
the observation of a fast baryon in the proton beam direction.
Detection of energetic neutrons indicate that isospin-1 exchanges 
are present, in particular charged pion exchange. 
Protons are sensitive to both isoscalar and isovector exchanges. 

Where the leading proton is close to the beam energy,
diffractive scattering is expected to be dominant.
Partonic scattering processes in such diffractive interactions give access to
quark-gluon configurations that are colour neutral but different from the well 
known hadrons.

The following measurements to obtain information on colour singlet exchange
are discussed here:
\begin{enumerate}
\item
the $ep$ structure function with a tagged baryon (Fig.~\ref{fig:feynman-lb}),
\item
the $ep$ structure function of diffractive exchange
(Figs.~\ref{fig:feynman-lb} ``$\pom$'' and ~\ref{fig:feynman-disrg}),
\item
di-jet and W-boson production in diffractive $\bar{p}p$ scattering
(Fig.~\ref{fig:feynman-tev}), and
\item
vector meson production in $ep$ interactions
(Fig.~\ref{fig:feynman-vm}).
\end{enumerate}

\subsection{\boldmath Tagged Baryon Production in $ep$ Collisions
\label{subsec:lb}}

\noindent
The production of protons and neutrons is studied in both the H1 and ZEUS 
experiments. 
A hadron calorimeter detects neutrons scattered at zero angle with respect 
to the proton direction. 
A series of Roman pot stations between the beam magnets serves as a proton 
spectrometer. 

\begin{figure}[hht]
\center
\epsfig{file=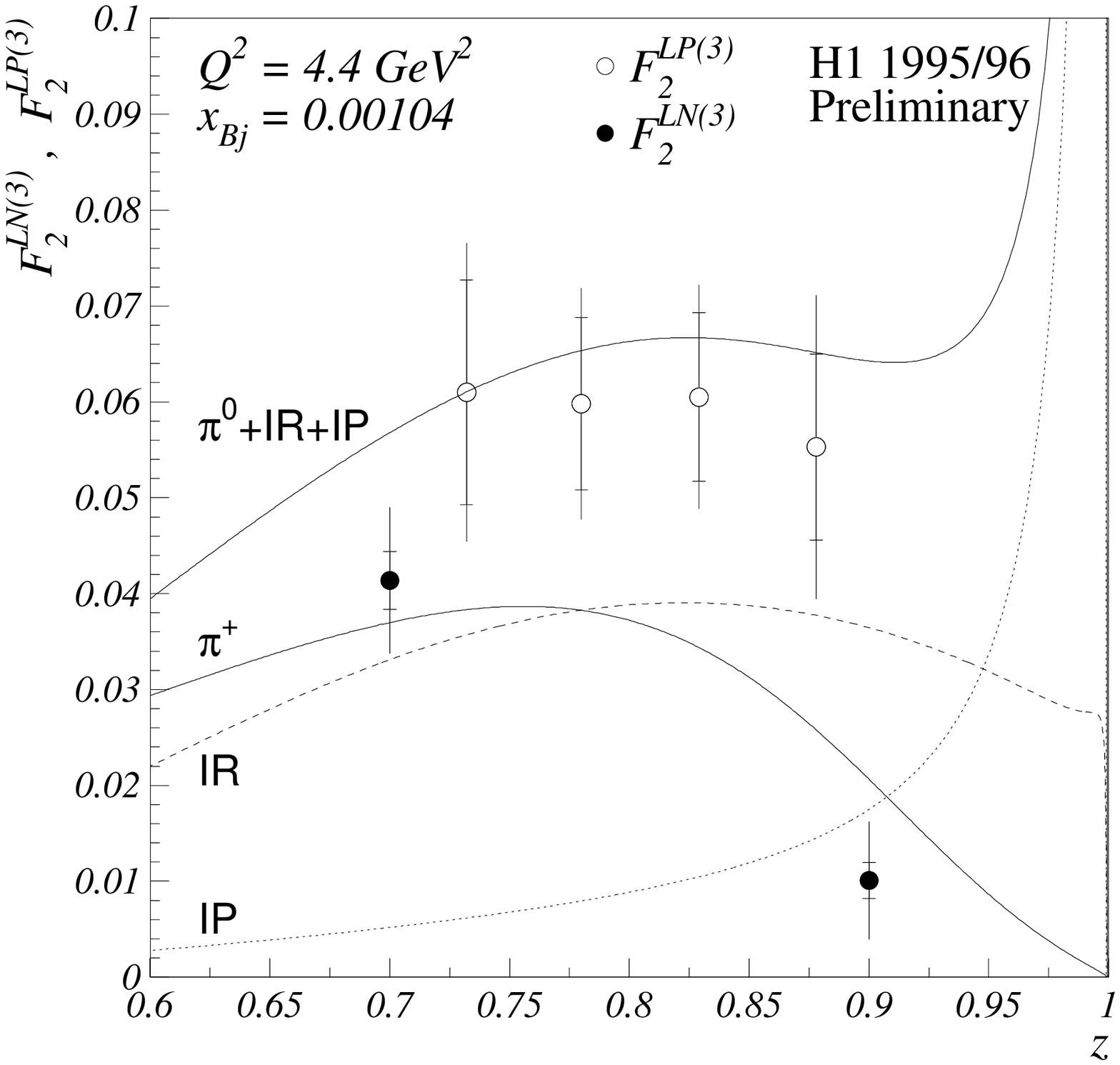,width=8cm}
\caption{ 
The $ep$ structure functions $F_2^{LB(3)}$ with a tagged proton 
(open circle) or neutron (full circle) are shown as a function 
of the baryon fractional energy $z$ at fixed photon virtuality 
$Q^2=4.4$ GeV$^2$ and parton fractional momentum 
$x_{Bj}=10^{-3}$ from H1 data.
The curves represent the prediction of a Regge model.  }
\label{fig:h1-lb}
\end{figure}

In Fig.~\ref{fig:h1-lb}, new measurements of the structure function 
from $ep$ collisions with a tagged baryon $F_2^{\rm LB(3)}$ 
are shown as a function of the baryon fractional energy 
$z= E_{lb}/E_p$ for fixed photon virtuality $Q^2=4.4$ GeV$^2$ and 
parton fractional momentum $x_{Bj}=10^{-3}$
(H1 collaboration ~\cite{h1-lb}). 
$F_2^{\rm LB(3)}$ was determined from cross section
measurements which were integrated over the
baryon transverse momenta in the range $0\le p_t\le 0.2$ GeV:
\begin{equation}
\frac{d^3\sigma}{dx_{Bj} dQ^2 dz} = 
\frac{2 \pi \alpha^2}{x_{Bj}\, Q^4} \left( 1 + ( 1 - y )^2 \right)
\, F_2^{\rm LB(3)}(x_{Bj},Q^2,z).
\end{equation}
Here $\alpha$ is the electro-magnetic coupling constant and $y$ denotes the 
inelasticity $y=Q^2/(x_{Bj} s_{ep})$.

\begin{figure}[hht]
\setlength{\unitlength}{1cm}
\begin{picture}(10.0,4.0)
\put(9.5,0.1)
{p,n}
\put(8.05,0.72)
{, $\reg$, $\pi$}
\put(6.5,0.72)
{$\xi =$}
\put(6.,0.)
{\epsfig{file=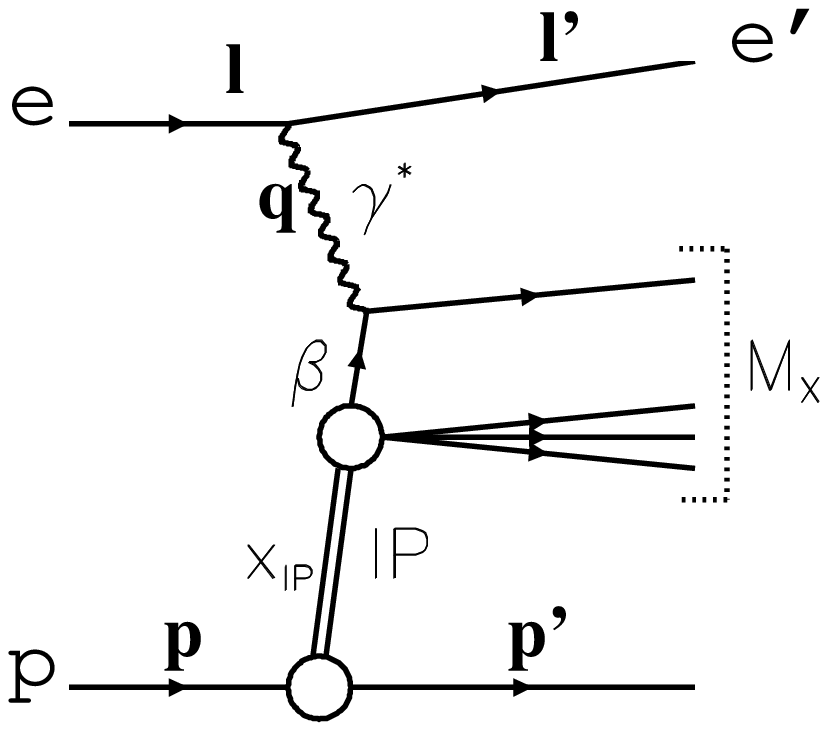,width=4cm}}
\end{picture}
\caption{A Feynman diagram of deep inelastic electron--proton
scattering with a tagged baryon is shown in the interpretation
of colour singlet exchange.
}
\label{fig:feynman-lb}
\end{figure}

The proton tagged structure function is found to be larger than that of the 
neutron tagged data. 
The curves are predictions of model calculations inspired by Regge 
phenomenology (Fig.~\ref{fig:feynman-lb}).  
In this picture, the proton data cannot be explained by $\pi^\circ$
exchange alone, since from the $\pi^\circ p$ and $\pi^+ n$ isospin $1/2$ 
states one would expect the proton measurement to be a factor two below the 
neutron data.
Instead, the proton data can be explained by an admixture of 
$\pi^\circ$, Reggeon ($f$, $\omega$) and Pomeron exchange.
The neutron data can be explained by charged pion exchange alone and
demonstrate the potential access to the pion structure function 
\cite{kopelyovich} in the new kinematic domain
at small parton momenta around $x\sim 10^{-3}$. 

\begin{figure}[hht]
\setlength{\unitlength}{1cm}
\begin{picture}(5.0,4.5)
\put(5.,4.)
{\epsfig{file=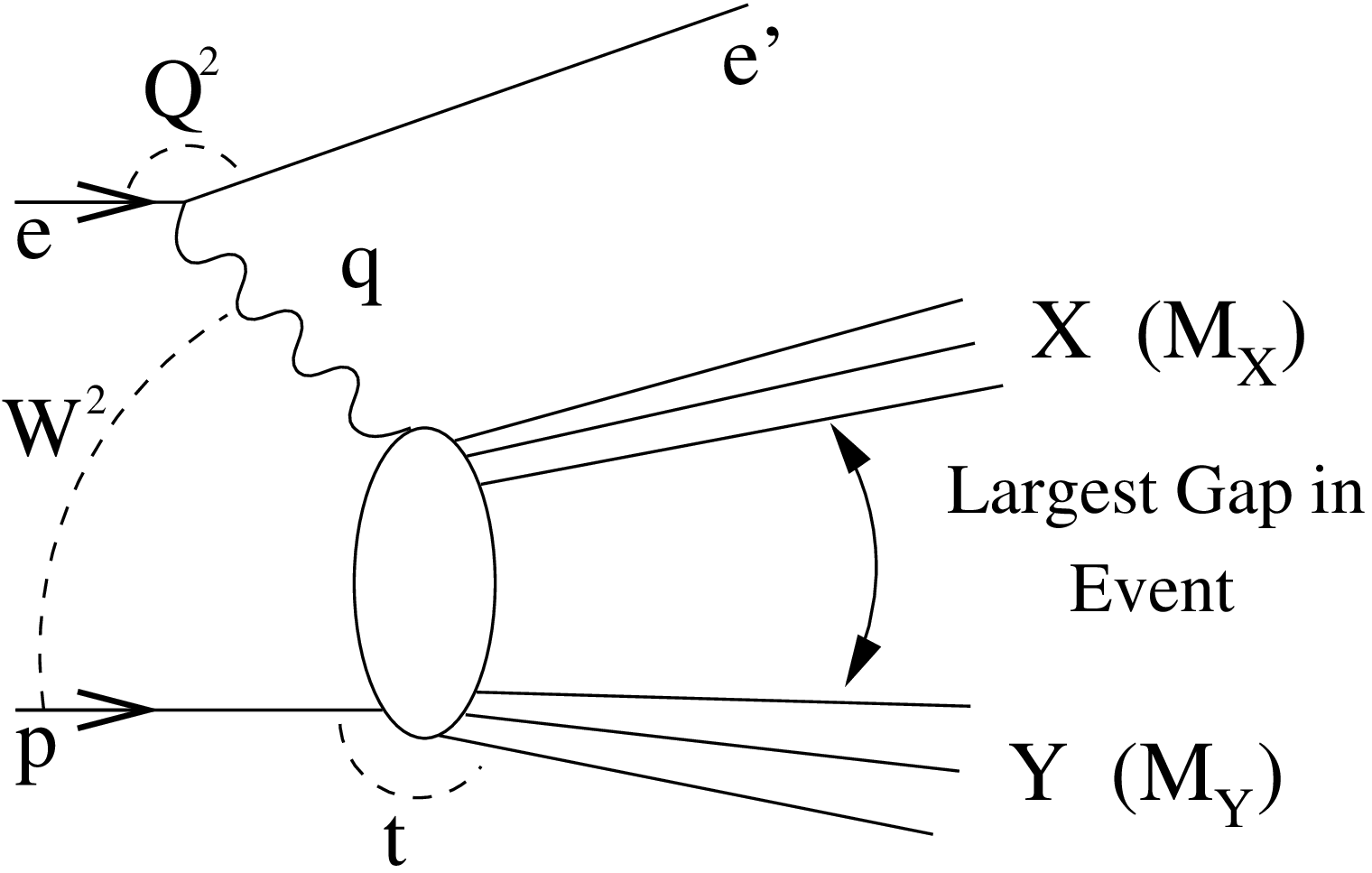,width=4cm,angle=270.0}}
\end{picture}
\caption{Feynman diagram of deep inelastic electron--proton
scattering with a large rapidity gap.
}
\label{fig:feynman-disrg}
\end{figure}

\begin{figure}[hht]
\center
\epsfig{file=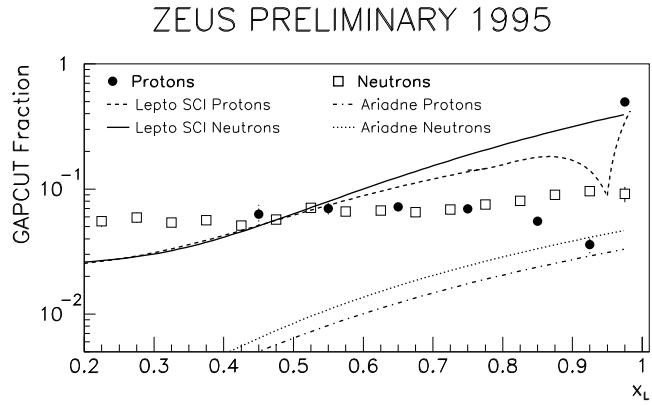,width=8.5cm}
\caption{ The rate of $ep$ collisions with a tagged proton (full circle) 
or neutron (open square) with simultaneous formation of a rapidity gap is 
shown as a function of the baryon fractional energy $x_{L}$
(ZEUS experiment).
The curves show the predictions of different Monte Carlo generator 
calculations.
}
\label{fig:zeus-lb}
\end{figure}

Further information on the type of the interaction process comes
from a new measurement of tagged baryons with 
the coincident formation of a large rapidity gap 
between the systems $X$ and $Y$ (Fig.~\ref{fig:feynman-disrg})
where $Y$ may or may not be observed in the main detector 
(ZEUS collaboration ~\cite{zeus-lb}).
In Fig.~\ref{fig:zeus-lb}, the rate of events with a large rapidity
gap is shown as a function of the baryon fractional energy 
$x_L\equiv z = E_{lb}/E_p$.
For $x_L \rightarrow 1$, the tagged proton production (full circle)
is dominated by diffractive processes.
For $x_L\ll 1$, the minimum gap size chosen for the analysis implies
that $M_Y>M_p$.
In this kinematic region, the rate of events with a large
rapidity gap is small and shows that
diffraction is not the main mechanism for the production of the baryons.

\subsection{The Partonic Structure of Diffractive Exchange
\label{subsec:cse}}

\noindent
Evidence for diffractive scattering processes in $ep$ interactions
can be obtained from different methods:
\begin{description}
\item[A]
tagging of highly energetic protons in the proton spectrometers
(Section~\ref{subsec:lb}),
\item[B]
from analysis of rapidity regions which are free of hadronic 
activity (``rapidity gap'', Fig.~\ref{fig:feynman-disrg}), or
\item[C]
from the mass distribution of the hadronic final state which is
observed in the main detector.
\end{description}

\subsection*{\boldmath The $ep$ Structure Function of Diffractive Exchange}

\begin{figure}[hht]
\center
\epsfig{file=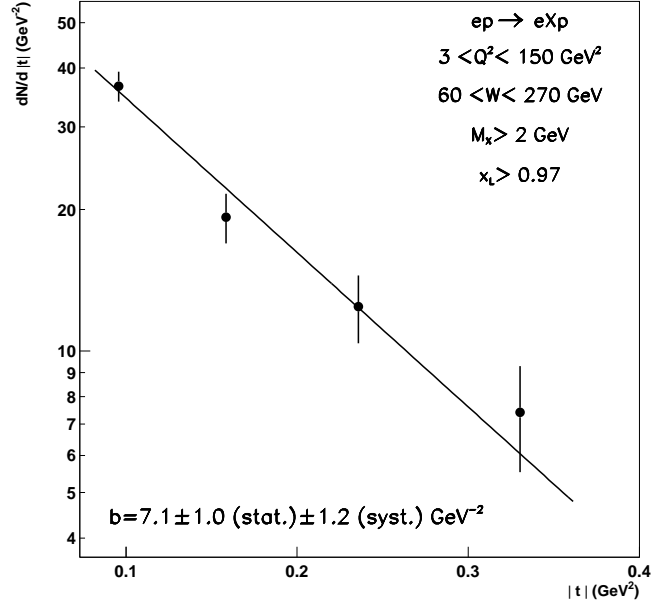,width=8.5cm}
\caption{ The distribution of the squared four-momentum transfer $\vert t\vert$ 
of diffractive $ep$ processes is shown together with an exponential fit 
to the data (ZEUS experiment).
}
\label{fig:zeus-t}
\end{figure}

\noindent
The structure function $F_2^{D(4)}$ for diffractive exchange 
\begin{equation}
\frac{d^4\sigma}{d\beta\, dQ^2 \, d\xi\, dt} =
\frac{2 \pi \alpha^2}{\beta \, Q^4} \left( 1 + ( 1 - y )^2 \right) \;
F_2^{D(4)}(\beta, Q^2, \xi, t)
\end{equation}
has been 
measured by the ZEUS collaboration using the tagged proton method A
~\cite{zeus-f2d4} as a function of the following four variables: 
\begin{enumerate}
\item
the virtuality $Q^2$ of the exchanged photon. 
\item
the squared four-momentum transfer $\vert t\vert = (p-p^\prime)^2$ 
from the proton side, 
\item
the momentum fraction 
$\xi = (Q^2+M_X^2)/(Q^2+s_{\gamma^* p})$ ($\xi \equiv x_\pom$), 
with $M_X$ being the mass of the diffractive system observed in the 
main detector, and
\item
the fractional momentum $\beta = x_{Bj}/\xi$.
\end{enumerate}
When interpreting this process in terms of the exchange 
of a colour singlet object (Fig.~\ref{fig:feynman-lb}), 
$\xi$ gives the fractional momentum that this object takes from the 
proton, and 
$\beta$ is the fractional momentum of the quark involved in the 
electron-quark scattering process.
Therefore, this deep inelastic scattering measurement gives access 
to the partonic structure of diffractive color singlet exchange and 
provides information on the corresponding $t$ distribution of this process.
A new measurement of the $t$ distribution is shown in 
Fig.~\ref{fig:zeus-t} (ZEUS collaboration \cite{zeus-lb}). 

The methods B (Fig.~\ref{fig:feynman-disrg}) and C of measuring the 
deep inelastic scattering of diffractive exchange 
have to integrate over some $t$-range and can here take advantage of the 
knowledge of the $t$ distribution of the proton tagged data. 
The two methods also do not include the detection of proton remnant
particles at small masses $M_Y$ of the dissociated proton system and integrate
over a small range of this mass (typically 1-4 GeV).
Since the acceptance of proton-tagged events is at the percent level, the 
statistics using methods B,C are much larger by far.

\begin{figure}[hht]
\setlength{\unitlength}{1cm}
\begin{picture}(16.,9.))
\put(0.,0.)
{\epsfig{file=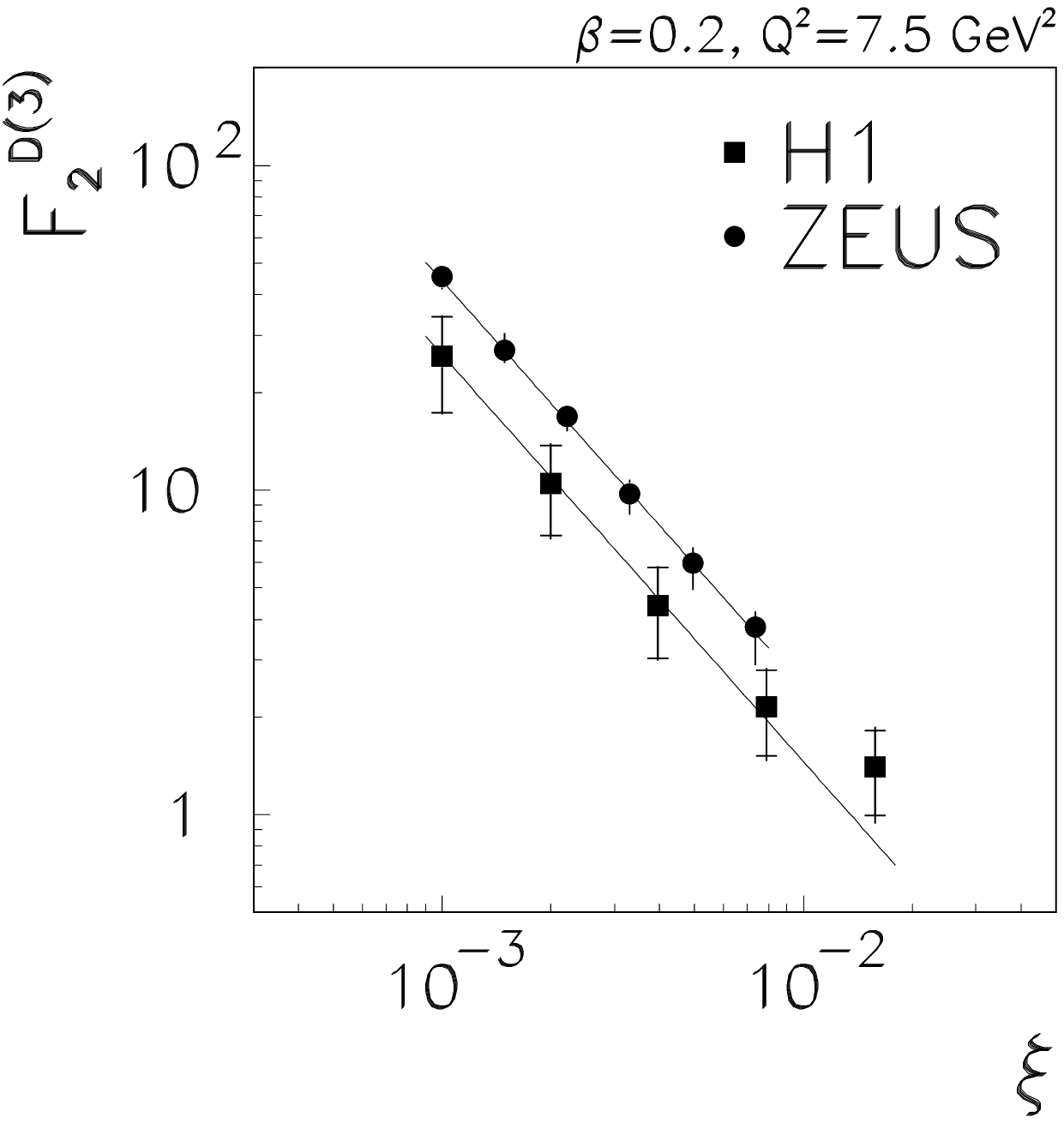,width=9cm}}
\put(7.3,0.)
{\epsfig{file=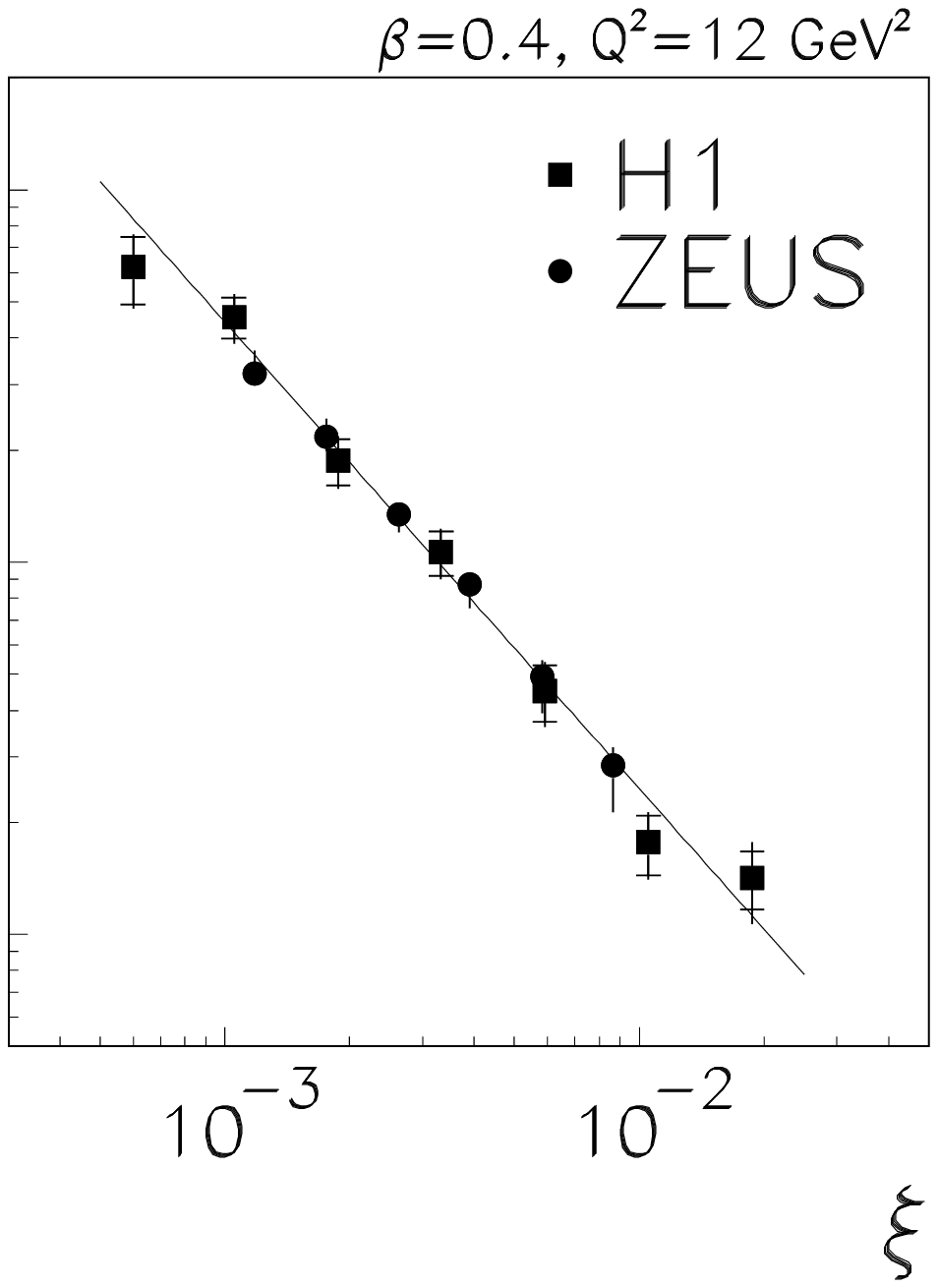,width=9cm}}
\end{picture}
\caption{ The $ep$ structure function $F_2^{D(3)}$ of diffractive exchange 
is shown as a function of the fractional energy $\xi$ of the exchanged 
object in two bins of the photon virtuality $Q^2$ and the parton 
fractional momentum $\beta$ (H1 
and ZEUS 
experiments). 
The curves have the functional form $\xi^{-n}$ and serve the guidance of 
the eye. }
\label{fig:f2d3}
\end{figure}

In Fig.~\ref{fig:f2d3}, a new triple differential structure 
function measurements $F_2^{D(3)}$ of the 
ZEUS collaboration (method C ~\cite{zeus-f2d3}) 
are compared with previous measurements by 
the H1 collaboration (method B ~\cite{h1-f2d3}).
The data are shown in a small selection of the large phase space covered 
as a function of the fractional momentum $\xi$, which the 
colour singlet object takes from the proton, in two bins of the parton 
momentum observable $\beta$ and the photon virtuality $Q^2$. 

At small $\xi$, they are consistent in these and surrounding phase space 
bins with a power law 
$\xi^{-n}$ and therefore are compatible with factorization
of the $\xi$ dependence.
The measured value by the ZEUS collaboration ~\cite{zeus-f2d3} is
$n-1=0.253\pm 0.017 ^{+0.077}_{-0.023}$ and is compatible with the
result of the H1 collaboration ~\cite{h1-f2d3}.
The measured value of $n$ is slightly larger than the value expected for 
soft Pomeron exchange in Regge inspired models ($n-1\sim 0.1$).

The $\beta$ and $Q^2$ dependence of $F_2^{D(3)}$ at fixed small value
of $\xi$ therefore gives the partonic structure of colour singlet exchange.  
The results of the two collaborations are consistent in most 
of the 15 phase space regions commonly covered, 
for example in Fig.~\ref{fig:f2d3}b, 
and call in a few of them for homework (Fig.~\ref{fig:f2d3}a), 
especially in an understanding of slightly different kinematic
regions covered in the squared momentum transfer $t$ and the mass
of the diffractive system $M_Y$.

\begin{figure}[hht]
\center\epsfig{file=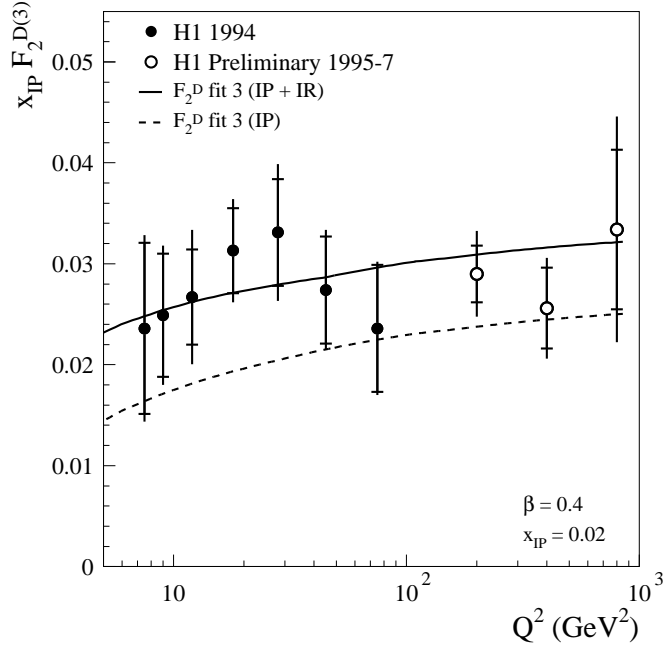,width=9.5cm}
\caption{The structure function $F_2$ of the diffractive 
exchange is shown as a function of the
virtuality $Q^2$ of the probing photon at 
the parton fractional momentum $\beta=0.4$ and at
the fractional momentum $\xi=x_{\pom}=0.02$ of the colour singlet 
object (H1 experiment).
The curves are QCD fits to measurements below $Q^2=100$ GeV$^2$.
}
\label{fig:f2-q2-d}
\end{figure}

In Fig.~\ref{fig:f2-q2-d}, the resolution scale $Q^2$ dependence of $F_2^{D(3)}$ of 
colour singlet exchange at large parton momenta $\beta=0.4$ is shown. 
This measurement has been newly extended to large $Q^2$ up to $800$ GeV$^2$
by the H1 collaboration ~\cite{h1-new-f2d3}. 
The data are at relatively large values of $\xi=0.02$ and can be described
by a dominant diffractive exchange (Pomeron exchange) together with
meson contributions (Reggeon exchange).

The $Q^2$ dependence of $F_2^{D(3)}$ 
is found to be consistent with flat 
which is very different from the structure function measurements of hadrons,
e.g., the proton structure function 
(Fig.~\ref{fig:f2-q2-p} ~\cite{h1-highq2}). 
It is also different from the $Q^2$ dependence of the photon structure 
function (Fig.~\ref{fig:f2-q2-g}). 
The different distributions can be understood from the QCD evolution equations:
the probability $f_q$ of finding a quark in the proton, color singlet exchange, 
or photon depends logarithmically on $Q^2$:                          
\begin{equation}
\frac{d f_q}{d \ln{Q^2}} = P_{qq} \otimes f_q + P_{qg} \otimes f_g + P_{q\gamma}
\label{eq:dglap}
\end{equation}
The $P_{ij}\otimes f_j$ denote the splitting functions convoluted with the
parton densities.
The first term $P_{qq} \otimes f_q$ represents the contribution of quarks
after radiating a gluon.
The second term $P_{qg} \otimes f_g$
gives the contributions of gluons that split into
a quark--anti-quark pair.
The third term $P_{q\gamma}$ adds the quarks resulting from the photon splitting
into a quark--anti-quark pair (relevant for photon only).

The proton structure function falls at large $x=0.4$ with increasing
resolution scale $Q^2$: 
the probability of finding a parton in the proton above the average
valence quark momentum decreases with increasing resolving power $Q^2$
(first term of eq.~(\ref{eq:dglap})).
The logarithmic increase of the photon structure function with $Q^2$
is caused by the third term of eq.~(\ref{eq:dglap}) which is to
first approximation independent of $Q^2$.

The structure function of diffractive exchange differs from those of the
proton and the photon:
the flat shape makes it distinct from a quark dominated object.
The large rate of diffractive exchange excludes an explanation by
photon exchange.
Instead, a large gluon density in the exchanged diffractive object 
can explain the observed $Q^2$ dependence of the structure function 
which is driven by the second term of eq.~(\ref{eq:dglap}).
Therefore the structure function measurement mainly probes the
gluon splitting into a quark--anti-quark pair and 
reflects the structure of the strong interactions.

This partonic structure of colour singlet exchange 
has been quantified by extracting
gluon and quark distributions from the diffractive
data using the structure function measurements alone (H1 Collaboration 
~\cite{h1-f2d3}) or in combination with jet cross section measurements
(ZEUS Collaboration ~\cite{zeus-diffpdf}).

Different final state observables have been measured in diffractive 
$ep$ scattering by the H1 and ZEUS Collaborations,
e.g., thrust ~\cite{h1-thrust,zeus-thrust},
di-jet cross sections ~\cite{h1-diffjet,zeus-diffpdf},
energy flow ~\cite{h1-eflow,zeus-thrust}, 
multiplicity ~\cite{h1-multi}, and
charm production ~\cite{h1-charm,zeus-charm}.
A large fraction of the measurements have been compared to Monte Carlo generators 
which simulate diffractive $ep$ scattering processes by the emission 
of colour singlet objects with the parton distributions as extracted 
from the fits to $F_2^{D(3)}$ mentioned above.
Overall, the data are well described by such simulations which demonstrates
a consistently working framework for understanding diffractive parton 
scattering processes in $ep$ collisions.
A deviation of this good description of the data may be seen in the
photoproduction of di-jets which is discussed below in the comparison
of the rates of diffractive processes at the HERA and Tevatron colliders.

Note that the picture of exchanging a colour 
singlet object with a partonic structure is not the only one to
describe the data:
interesting alternative approaches exist which need fewer
parameters and describe certain aspects of the data well.
Examples are electron scattering off a quark or a gluon of the proton
with colour neutralization by the exchange 
of a second parton that cancels the colour charge, or
models that predict the $\beta$ dependence of $F_2^{D(3)}$, or
the concept of fracture functions ~\cite{sassot}.
For reviews of the different approaches refer to, e.g., ~\cite{heraws96,jung}.

\subsection*{Colour Singlet Exchange at the Tevatron}

\noindent
The methods used by the Tevatron experiments CDF and D0 to select
diffractive scattering processes are detection of leading protons (method A) or
measurement of rapidity gaps (method B).
The observables used to analyse the diffractive exchange
are di-jet formation and the production of W-bosons.

\begin{figure}[hht]
\setlength{\unitlength}{1cm}
\begin{picture}(5.0,6.0)
\put(-0.8,0.)
{\epsfig{file=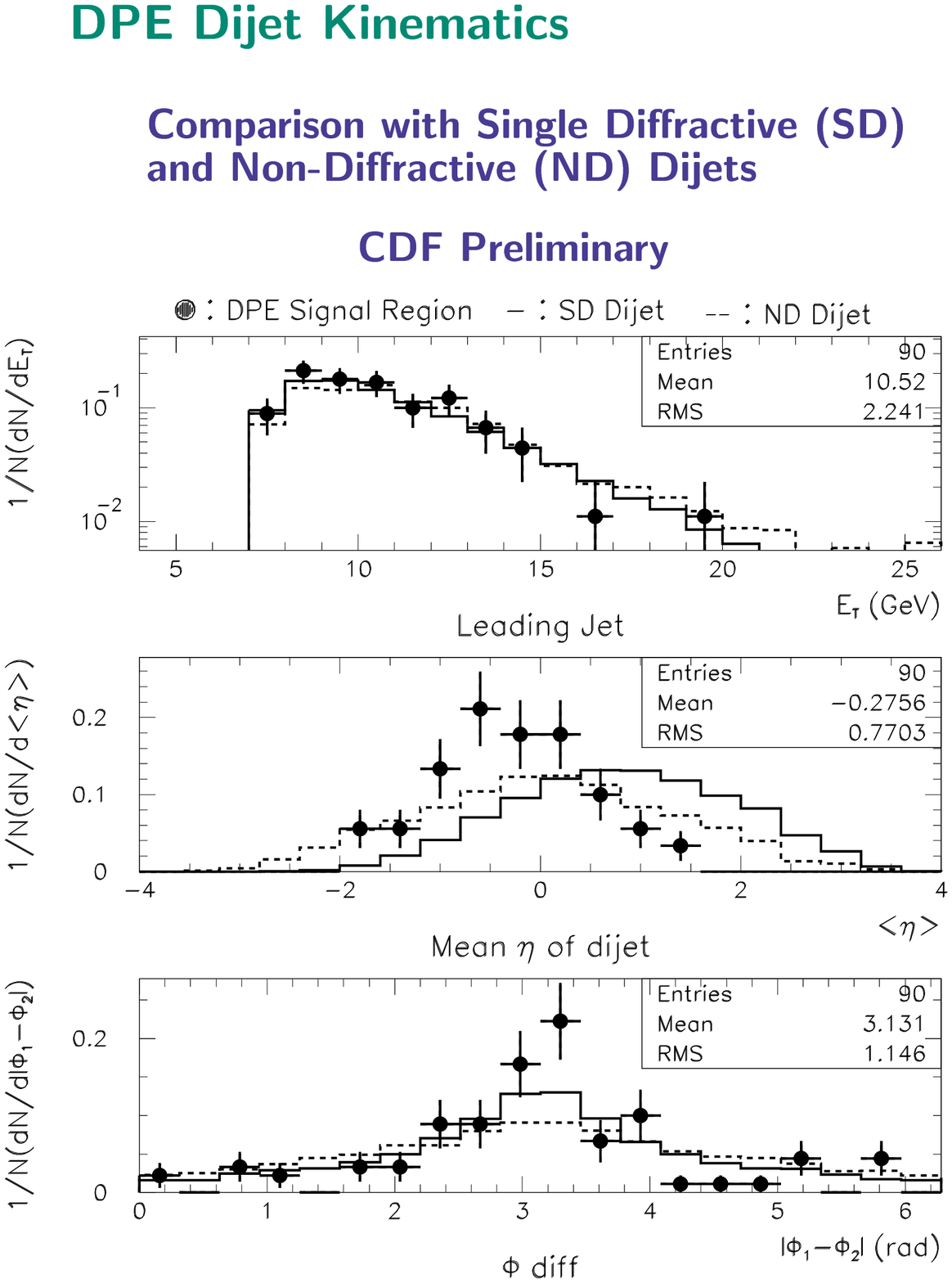,bbllx=0,bblly=400,bburx=525,bbury=612,width=13cm,clip=}}
\end{picture}
\caption{ The shape of the transverse energy spectrum of the leading jet is 
shown from non-diffractive (dashed histogram), single diffractive 
(full histogram), and double diffractive (full circles) events in $\bar{p}p$ 
collisions (CDF experiment).
}
\label{fig:cdf-diff}
\end{figure}

Both experiments have observed events involving the exchange of one
or - as a new result - two colour singlet objects 
(Fig.~\ref{fig:feynman-tev}a,c ~\cite{cdf-dijet,d0-dijet}).
In the latter process, the jets are produced centrally and 
in each beam direction a large rapidity gap 
or a tagged proton is observed.
In Fig.~\ref{fig:cdf-diff}, the shapes of transverse energy $E_t^{jet}$
distributions of the leading jets in di-jet events are compared for 
single and double colour singlet exchange and non-diffractive data
(CDF Collaboration).
The $E_t^{jet}$ range covered and the similarity of these distributions 
give several interesting observations: 
\begin{figure}[hht]
\center\epsfig{file=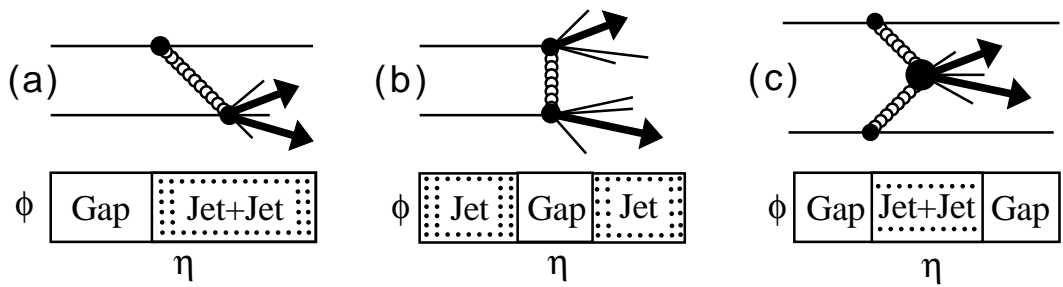,width=14cm}
\caption{Di-jet production at the Tevatron with 
a) rapidity gap on one beam side, 
b) rapidity gap between the jets, or
c) central jet production with two rapidity gaps.}
\label{fig:feynman-tev}
\end{figure}

The diffractive di-jet production results from the same type of 
parton--parton scattering processes as the non-diffractive data. 
In the latter case, the fractional momenta of the partons from the proton are 
small $x \sim E_t^{jet}/\sqrt{s} \sim 10/1800$ and therefore
likely to come from gluon--gluon scattering processes.
In the diffractive case with the exchange of one or two 
colour singlet objects, the center-of-mass energy of the scattering process
is much smaller than that of the 
$\bar{p}p$ beams since these objects carry only a fraction
of the beam proton energy.
Nevertheless, the jet transverse energy reaches out to 
$E_t^{jet}\sim 20$ GeV such that almost the full energy of these objects
is involved in the hard parton--parton scattering process.

\begin{figure}[hht]
\center
\epsfig{file=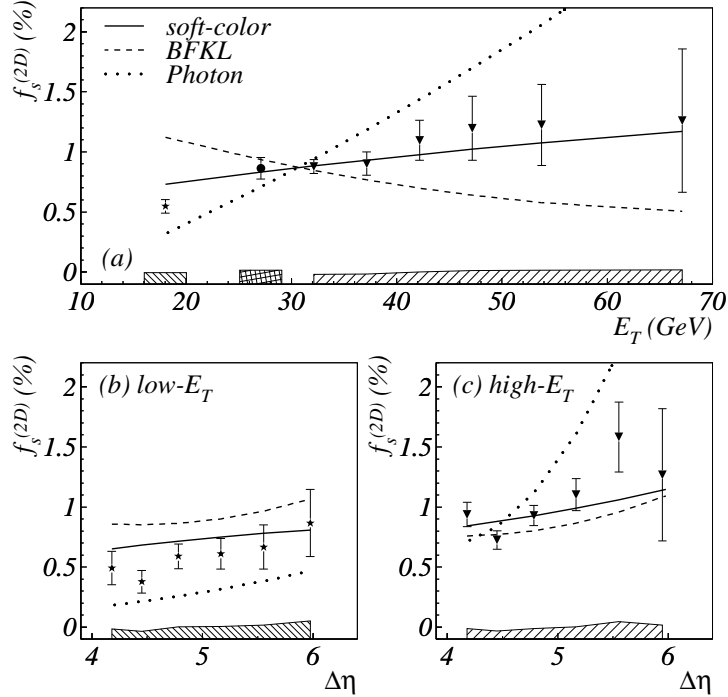,width=10.5cm}
\caption{
The rate of $\bar{p}p$ events with a rapidity gap between two jets is shown 
from preliminary D0 data:
a) differentially in the transverse jet energy $E_t^{jet}$, and in
b,c) as a function of the gap size $\Delta\eta$ for low and high values of 
$E_t^{jet}$. 
The curves represent the predictions of different model calculations. }
\label{fig:d0}
\end{figure}

Both Tevatron experiments have observed events with a rapidity gap
between two jets (Fig.~\ref{fig:feynman-tev}b ~\cite{cdf-dijet,d0-hcse}).
In these events, the full energy of the exchanged object is involved 
in the jet production process
and the object is probed at very large squared four-momentum 
transfer $\vert t \vert$ of the order of $(E_t^{jet})^2$.
In Fig.~\ref{fig:d0}, the rate of events with such a colour singlet
exchange relative to non-diffractive events is shown from the
D0 collaboration.
The distribution of the size of the rapidity gap is shown to be 
within errors independent of the jet transverse energy 
(Fig.~\ref{fig:d0}b,c).
In Fig.~\ref{fig:d0}a, the rate is given as a function of the jet 
transverse energy which has a tendency to rise with increasing 
$E_t^{jet}$.

The data are sufficiently precise to discriminate different models
of colour singlet exchange:
they exclude the exchange of a photon (dotted curves in Fig.~\ref{fig:d0})
and a calculation using two hard gluons (``BFKL'', dashed curves ~\cite{bfkl}).
The data can be consistently described by a model calculating the 
exchange of one energetic gluon with an additional parton to
ensure colour neutrality (full curve ~\cite{softcolour}).

\begin{figure}[hht]
\center
\epsfig{file=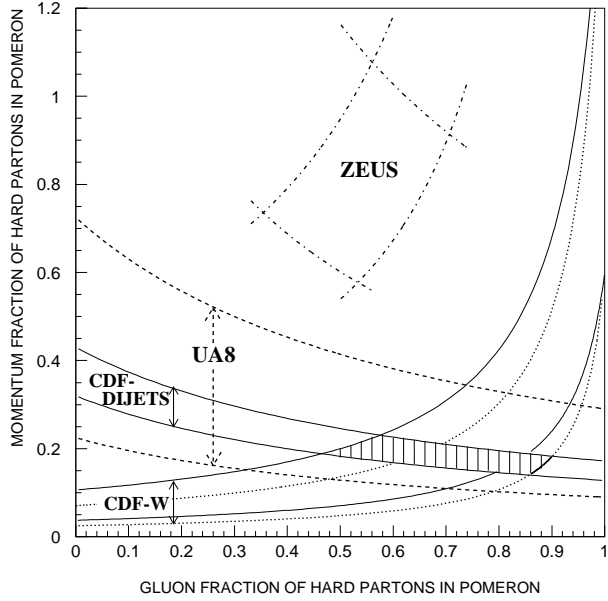,width=8cm}
\caption{
The total momentum fraction carried by the partons of the diffractive exchange 
object is shown as a function of the relative gluon contribution. 
The lower bands represent the region allowed from the measurements of diffractive 
di-jet production in 
$\bar{p}p$ collisions and $W$ production respectively 
(CDF experiment).
The bands in the upper half of the figure are a comparison of diffractive 
jet production with diffractive structure function measurements in $ep$ collisions
(ZEUS experiment). 
}
\label{fig:cdf-gluon}
\end{figure}

The CDF experiment has observed the production of W-bosons
in diffractive scattering processes ~\cite{cdf-w}.
These events have essentially one lepton and missing transverse 
energy and can be interpreted as resulting from quark--anti-quark fusion.
A comparison of the diffractive W-boson rate with that of the di-jet 
production is shown in Fig.~\ref{fig:cdf-gluon} as a function of the
relative gluon contribution in the colour singlet object and is
expressed as a momentum sum rule.
The gluon contribution is found to be large $0.7\pm 0.2$
which is well compatible with previous (shown in the figure) 
and new fits of the ZEUS 
collaboration ~\cite{zeus-diffpdf} (not shown) and previous results 
of the H1 collaboration ~\cite{h1-f2d3}.

\subsection*{Different Rates of Diffractive Processes at the
HERA and Tevatron Colliders}

\noindent
While the large gluon component is consistently observed in 
diffractive processes at HERA and the Tevatron, the rate of such
events is found to be largely different:
at HERA, the rate of diffractive deep inelastic scattering events
is of the order of $10\%$.
In the phase space regions covered so far,
the HERA final state data are overall consistently described 
when compared to calculations that
use the parton distributions resulting from the diffractive 
structure function measurements.
Using the same parton distributions for the Tevatron diffractive
data, the predicted rate is much larger than the observed rate
of the order of $1\%$ ~\cite{cdf-w,whitmore}.
This discrepancy can, e.g., be expressed in terms of a momentum
sum rule as shown in Fig.~\ref{fig:cdf-gluon}.
The inconsistency is a puzzle which is under lively discussion.

\begin{figure}[hht]
\center
\epsfig{file=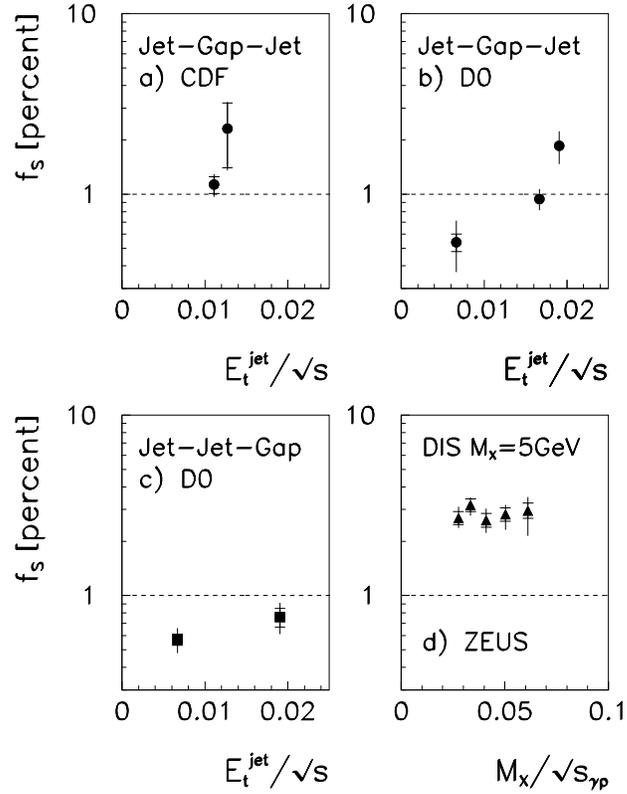,width=9cm}
\caption{
A comparison of the measured rates of diffractive interactions is shown as a 
function of the energy involved in the interaction relative to
the total hadronic center of mass energy:
a,b) ~rapidity gap signature between two jets of transverse energy
above $E_t^{jet}$ from $\bar{p}p$ collisions
(CDF 
and D0 
experiments),
c) ~diffractive di-jet production in $\bar{p}p$ collisions
(D0 data), and
d) ~diffractive structure function in $ep$ collisions where $M_X$ denotes 
the mass of the diffractive system observed in the main detector
(ZEUS experiment).
}
\label{fig:hera-tev}
\end{figure}

Instructive measurements have been made, allowing the 
energy $E_{ia}$ involved in the interaction to be measured
relative to the total hadronic center-of-mass energy $\sqrt{s}$.
In Fig.~\ref{fig:hera-tev}, rates of diffractive events are shown as
a function of the ratio $E_{ia}/\sqrt{s}$.

For the Tevatron jet results ~\cite{cdf-dijet,d0-dijet,d0-hcse}, 
the jet transverse energy $E_t^{jet}$
at the threshold has been used as a measure of $E_{ia}$
(Fig.~\ref{fig:hera-tev}a-c).
The rate of diffractive events appears to decrease as the 
total center-of-mass energy $\sqrt{s_{\bar{p}p}}$ becomes 
large relative to the energy involved in the scattering process.

Such dependence can, e.g., be explained by the increased potential
of destroying the rapidity gap by beam remnant interactions
which may be formulated in a reduced survival probability for the
rapidity gap.
Different other explanations have been suggested,
key words are here absorption corrections, 
flux renormalization, or other means of factorization breaking 
~\cite{glm,dino,schlein}.

For the HERA data, two measurements are discussed here:
in the case of deep inelastic scattering data, the mass $M_X$ of the 
diffractive system has been taken as a measure of $E_{ia}$.
The data in Fig.~\ref{fig:hera-tev}d
are consistent with being flat as a function of $M_X/\sqrt{s_{\gamma^* p}}$
~\cite{zeus-f2d3} and show no indication of a decreasing survival
probability.

\begin{figure}[hht]
\center
\epsfig{file=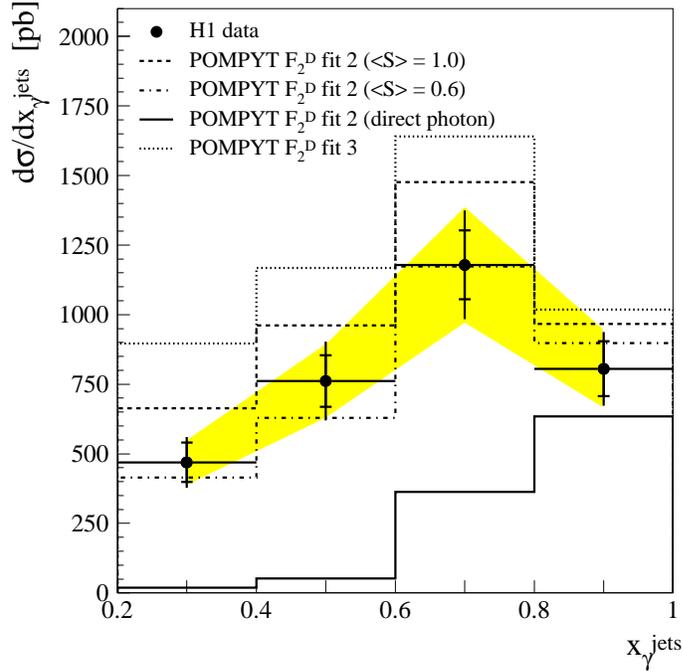,width=10cm}
\caption{Photoproduction of di-jets in diffractive $ep$
collisions is shown as a function of the fractional momentum $x$ 
of the parton from the photon (H1 data).
The dotted and dashed histograms show Monte Carlo generator 
calculations using two different parton distribution functions 
for the colour singlet exchange which were extracted from structure 
function measurements.
The full histogram represents the contribution of 
direct photon--proton processes in diffractive interactions.
In the dash-dotted histogram, a ``rapidity gap survival factor'' of $S=0.6$ 
was applied to the calculation for the generated momenta 
of the partons from the photon below $x=0.8$.
}
\label{fig:h1-diffjet}
\end{figure}

In photoproduction of di-jets, the fractional momentum $x$ of the
parton from the photon is related to the ratio 
$E_t^{jet}/\sqrt{s_{\gamma p}}$.
In Fig.~\ref{fig:h1-diffjet}, the di-jet cross section from
diffractive scattering processes is shown as a function of $x$ 
from H1 data ~\cite{h1-diffjet}.
At large $x\sim 1$, where the direct photon contribution dominates,
the data are described by the calculations 
of the POMPYT generator ~\cite{pompyt} when using 
the parton distribution functions for the colour singlet 
exchange as extracted from the $F_2^{D(3)}$ measurements.
However, at $x<0.8$ the data are better described, if an overall
reduction factor of $S=0.6$ is applied to the calculation of the
resolved photon--proton interactions.
This observation hints for a reduced survival probability of the
rapidity gap in resolved photon--proton processes.
Owing to the presence of a proton and a photon remnant,
these $\gamma p$ processes are similar to that of diffractive processes in
$\bar{p}p$ collisions.
In the future, more extended and precise measurements of the photoproduction 
of jets may help in the understanding of the different diffractive 
rates observed by the HERA and Tevatron experiments.


\subsection{\boldmath Vector Meson Production in $ep$ Collisions}

\begin{figure}[hht]
\center\epsfig{file=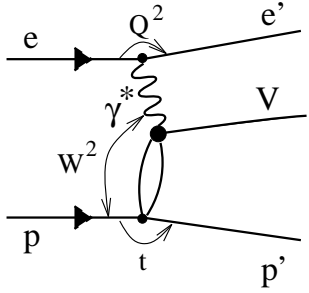,width=4.0cm}
\caption{Feynman diagram of vector meson production in electron--proton
scattering.
}
\label{fig:feynman-vm}
\end{figure}

\noindent
In elastic vector meson production from $ep$ collisions,
the full energy of the colour singlet object is involved
in the scattering process (Fig.~\ref{fig:feynman-vm}).
Of special interest are processes with a hard scale such as
\begin{enumerate}
\item
the mass $M_V$ of a heavy vector meson, 
\item
the virtuality $Q^2$ 
of the photon in a deep inelastic scattering process, or 
\item
the squared four-momentum transfer $t$ of the colour singlet
exchange.
\end{enumerate}
Such processes allow perturbative QCD calculations to be
compared with the measurements and therefore give additional
information on colour singlet exchange as well as on the
proton and the vector meson \cite{crittenden}.

In this context, the following measurements of vector meson 
production in $ep$ collisions at HERA are discussed here:
\begin{enumerate}
\item
vector meson cross sections and their dependencies on the center-of-mass
energy $\sqrt{s_{\gamma p}}$, the photon virtuality $Q^2$, and the
squared momentum transfer $t$, and 
\item
photoproduction of $J/\psi$ mesons from proton and nuclear targets.
\end{enumerate}

\subsection*{Measurements Related to the Gluon Distribution of the Proton}

\begin{figure}[hht]
\center
\epsfig{file=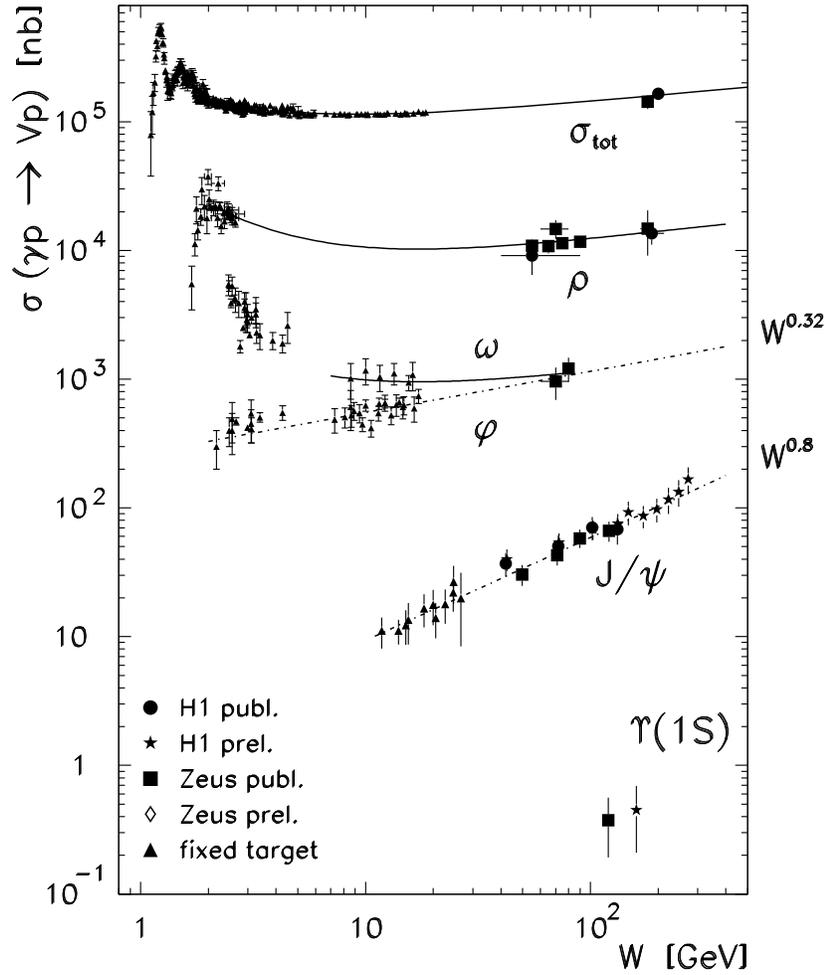,width=11cm}
\caption{
The photon-proton total cross section and vector meson cross sections from 
fixed target experiments and $ep$ collisions at HERA
are shown as a function of the photon-proton center of mass energy 
$W \equiv \sqrt{s_{\gamma p}}$. 
The full curves represent the predictions of a Regge model. 
The dashed curves are functions of the form 
$W^\delta\equiv (s_{\gamma p})^{2\epsilon}$ to guide the eye. }
\label{fig:sigma-vm}
\end{figure}

\noindent
In Fig.~\ref{fig:sigma-vm}, a compilation of the measurements of 
the total photoproduction cross section $\sigma_{\gamma p}$ and elastic
vector meson cross sections $\sigma_{\gamma p}^{V}$ up to the production
of $\Upsilon$ ~\cite{zeus-ups,h1-ups} is shown as a 
function of the photon--proton center-of-mass energy 
$W\equiv \sqrt{s_{\gamma p}}$.
The measured total cross section is at large center-of-mass energies
compatible with a slowly rising distribution as 
\begin{equation}
\sigma_{\gamma p}\sim s^\epsilon
\end{equation}
with $\epsilon\sim 0.095\pm 0.002$ ~\cite{epsilon}.
The optical theorem relates the total cross section to the 
imaginary part of the amplitude of forward elastic scattering.
Therefore, elastic vector meson cross sections should rise 
with approximately twice the power:
\begin{equation}
\sigma_{\gamma p}^{V}\sim s^{2\epsilon} \; .
\label{eq:sigma-vm}
\end{equation}
Photoproduction of light vector mesons ($\rho$, $\omega$, $\varphi$)
show an increase in the production that is compatible with this 
prediction.
However, photoproduction of the heavy $J/\psi$ mesons exhibit a stronger
dependence on the center-of-mass energy with $\epsilon\sim 0.2$.

\begin{figure}[hht]
\setlength{\unitlength}{1cm}
\begin{picture}(5.0,9.0)
\put(10.5,8.1)
{b)}
\put(5.5,8.1)
{a)}
\put(0.0,0.0)
{\epsfig{file=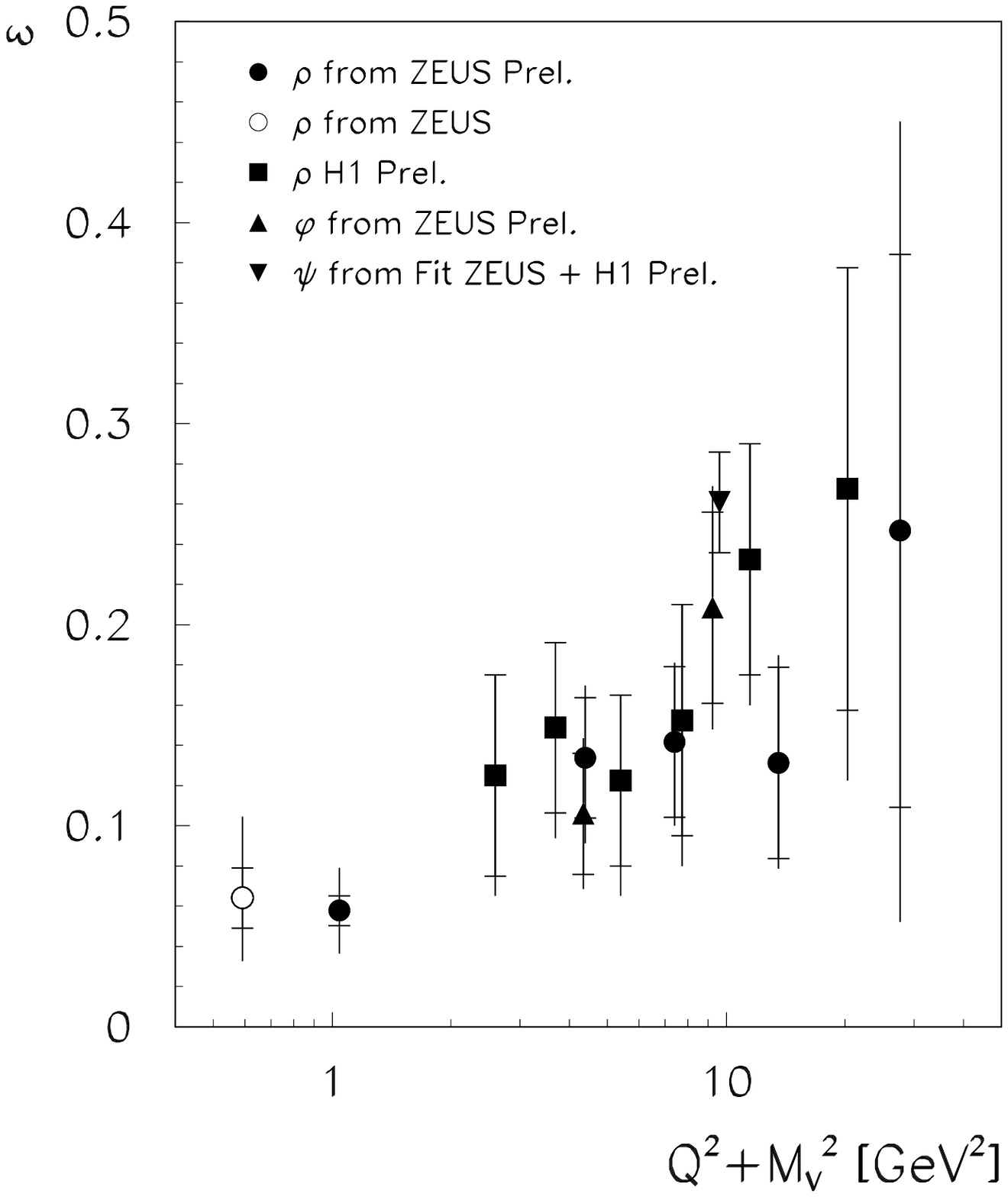,width=8.5cm}}
\put(8.0,0.0)
{\epsfig{file=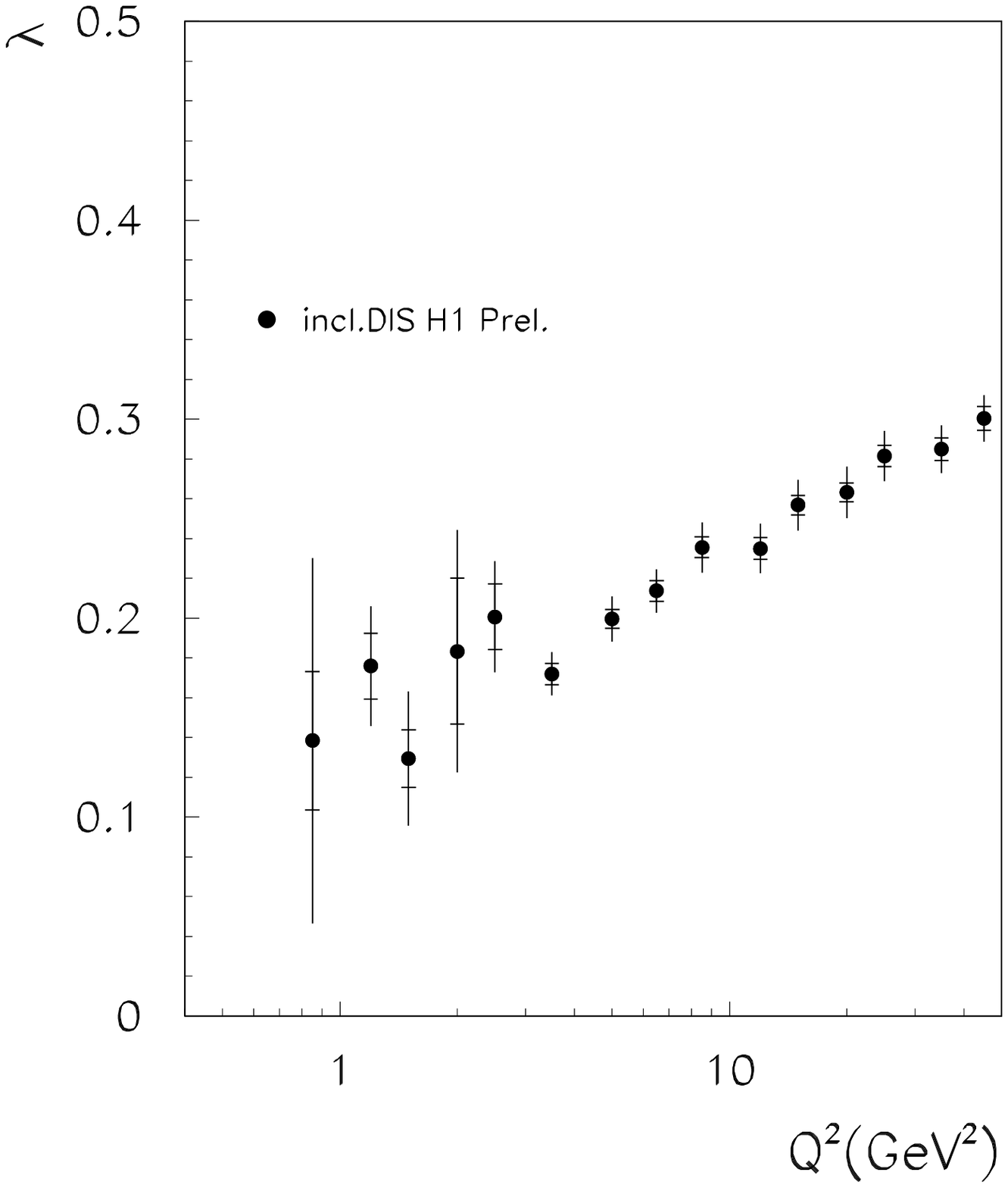,width=8.5cm}}
\end{picture}
\caption{
a) The energy dependence of vector meson cross sections
$\sigma_{\gamma p}^V \sim (s_{\gamma^* p})^{2\epsilon}$ 
in $ep$ collisions is shown as a 
function of the scale taken here to be the sum of the photon virtuality 
$Q^2$ and the vector meson squared mass  $M_ V^2$ (H1 and ZEUS experiments). 
b) ~The energy dependence of inclusive cross section measurements 
$\sigma_{\gamma^*p} \sim (s_{\gamma^* p})^\lambda$  is shown as a function 
of the photon virtuality $Q^2$ from H1 data. }
\label{fig:lambda}
\end{figure}

A steeper energy dependence is also observed for light vector meson
production in deep inelastic scattering processes:
in Fig.~\ref{fig:lambda}a, the energy dependence of 
new $\sigma_{\gamma p}^{V}$ measurements by the H1 and ZEUS Collaborations
~\cite{h1-rho,zeus-rho,zeus-phi,h1-psi}
was again expressed in terms of the fit parameter $\epsilon$ using
eq.~(\ref{eq:sigma-vm}) and is shown as a function of the scale.
The scale was here chosen to be the sum of the photon virtuality 
and the vector meson squared mass $Q^2+M_{V}^2$.
The parameter $\epsilon$ is found to increase with increasing scale.

Such energy dependence is similar to that observed in inclusive
deep inelastic scattering cross sections (Fig.~\ref{fig:lambda}b
~\cite{h1-f2-534}).
At fixed $Q^2$ and small parton momenta $x_{Bj}$, the total 
photon--proton cross section $\sigma_{\gamma^* p}$
is directly related to the large
gluon density observed in the proton which gives rise to the 
$(x_{Bj})^{-\lambda}$ dependence of the proton structure function $F_2$.
Using the relation $x_{Bj} s_{\gamma^* p} \approx Q^2 = const.$ 
gives an energy dependence of the cross section as
$\sigma_{\gamma^*p}\sim F_2\sim (x_{Bj})^{-\lambda}\sim (s_{\gamma^*p})^\lambda$.
The similar energy dependencies observed in vector meson production 
(Fig.~\ref{fig:lambda}a) and
inclusive deep inelastic scattering processes (Fig.~\ref{fig:lambda}b) 
is suggestive of 
sensitivity of the vector meson data to the gluon distribution of the
proton.

\begin{figure}[hht]
\center
\epsfig{file=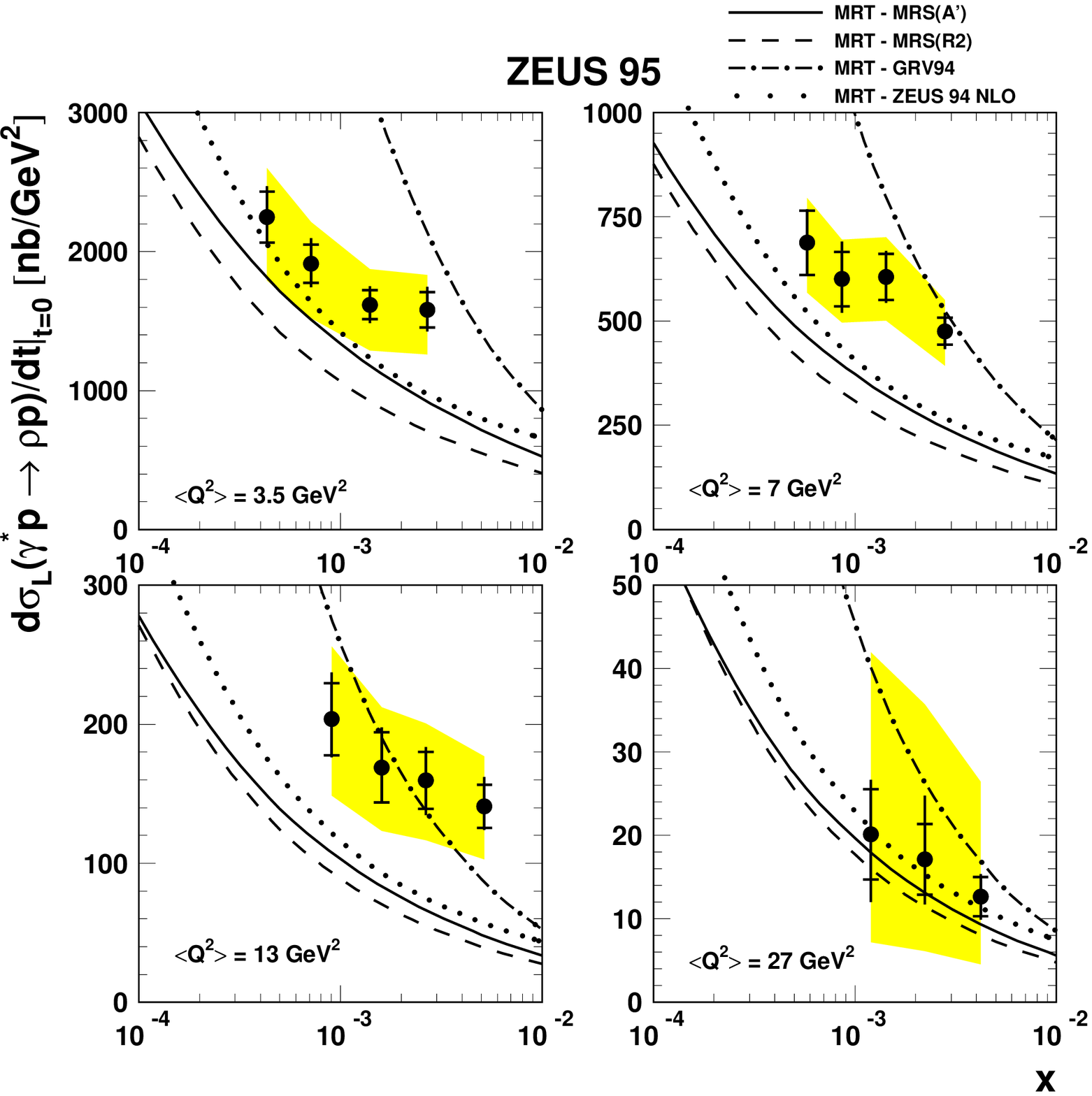,width=12cm}
\caption{
The longitudinal $\rho$ meson cross section from deep inelastic $ep$ scattering 
is shown as a function of the proton parton fractional momentum $x_{Bj}$ in four 
bins of the photon virtuality $Q^2$ (ZEUS experiment). 
The curves represent the predictions of QCD calculations using different gluon distributions of the proton. }
\label{fig:rho}
\end{figure}

\begin{figure}[hht]
\center
\epsfig{file=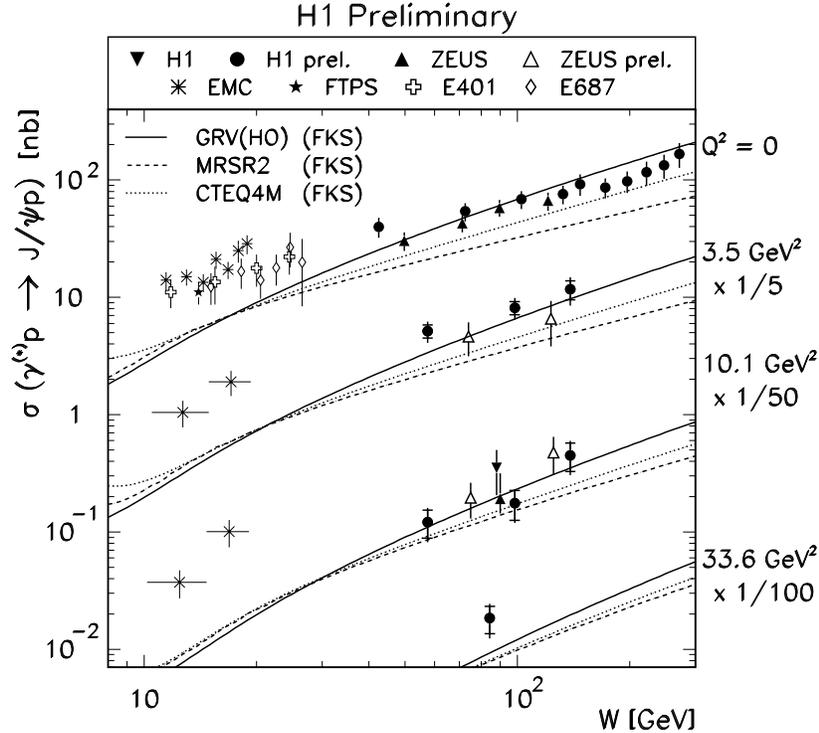,width=11cm}
\caption{ 
The $J/\psi$ cross section from fixed target experiments and $ep$ collisions 
at HERA is shown as a function of 
the photon-proton center of mass energy $W$ in four bins of the photon 
virtuality $Q^2$. 
The curves represent the predictions of QCD calculations using different 
gluon distributions of the proton. 
}
\label{fig:psi}
\end{figure}

In Fig.~\ref{fig:rho}, the longitudinal component of the $\rho$ meson
production cross section is shown as a function of $x_{Bj}$ in 
four bins of $Q^2$ ~\cite{zeus-rho}.
Similarly, the $J/\psi$ production cross section is shown in 
Fig.~\ref{fig:psi} as a function of the photon--proton center-of-mass
energy $W\equiv \sqrt{s_{\gamma^*p}}$ ~\cite{h1-psi}.

\begin{figure}[hht]
\center\epsfig{file=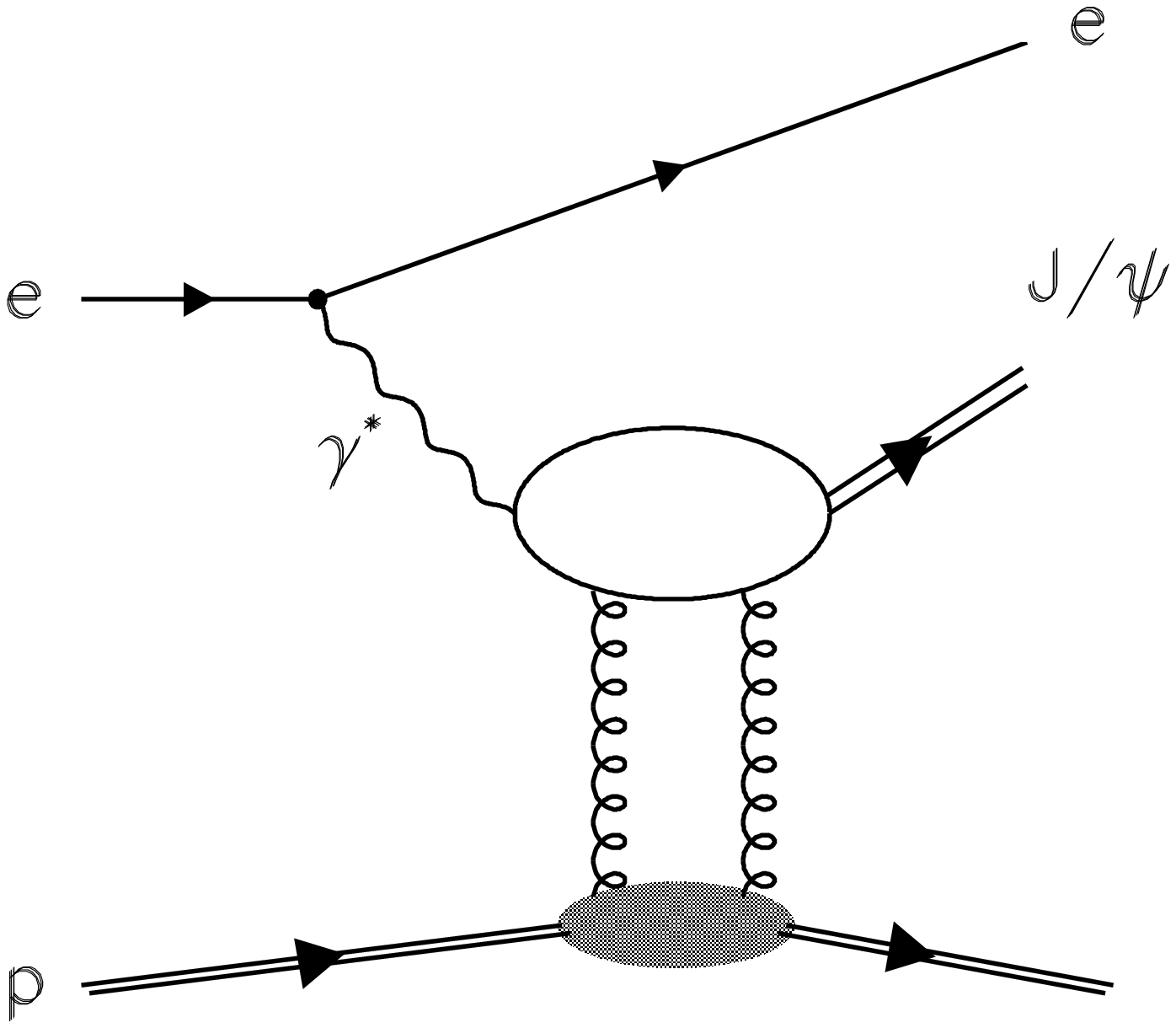,width=4cm}
\caption{Feynman diagram of $J/\psi$ meson production
in electron--proton scattering with two-gluon exchange.
}
\label{fig:feynman-psi}
\end{figure}

The measurements can be described by perturbative QCD calculations 
which use existing parameterizations of the gluon distributions in the 
proton (curves ~\cite{mrt,fks}).
The calculations use the square of the gluon density to account for
the colour neutrality of the exchanged object 
(e.g. Fig.~\ref{fig:feynman-psi}).
Therefore, the comparisons of the data with the calculations give
a highly sensitive measure of the gluons in the proton.
A further component of the calculations is the mechanism for 
formation of the vector 
meson such that the comparisons to the data will give new information
also on this part of the process.

\subsection*{Measurements Related to Quark--Anti-Quark States}

\noindent
In elastic vector meson production, the squared momentum transfer $t$,
which is exchanged between the vector meson and the proton, 
gives information on the size of the interaction region.
Such $t$ distributions can be fitted for small values of $t$ using
an exponential distribution $\exp{(-bt)}$. 

\begin{figure}[hht]
\center
\epsfig{file=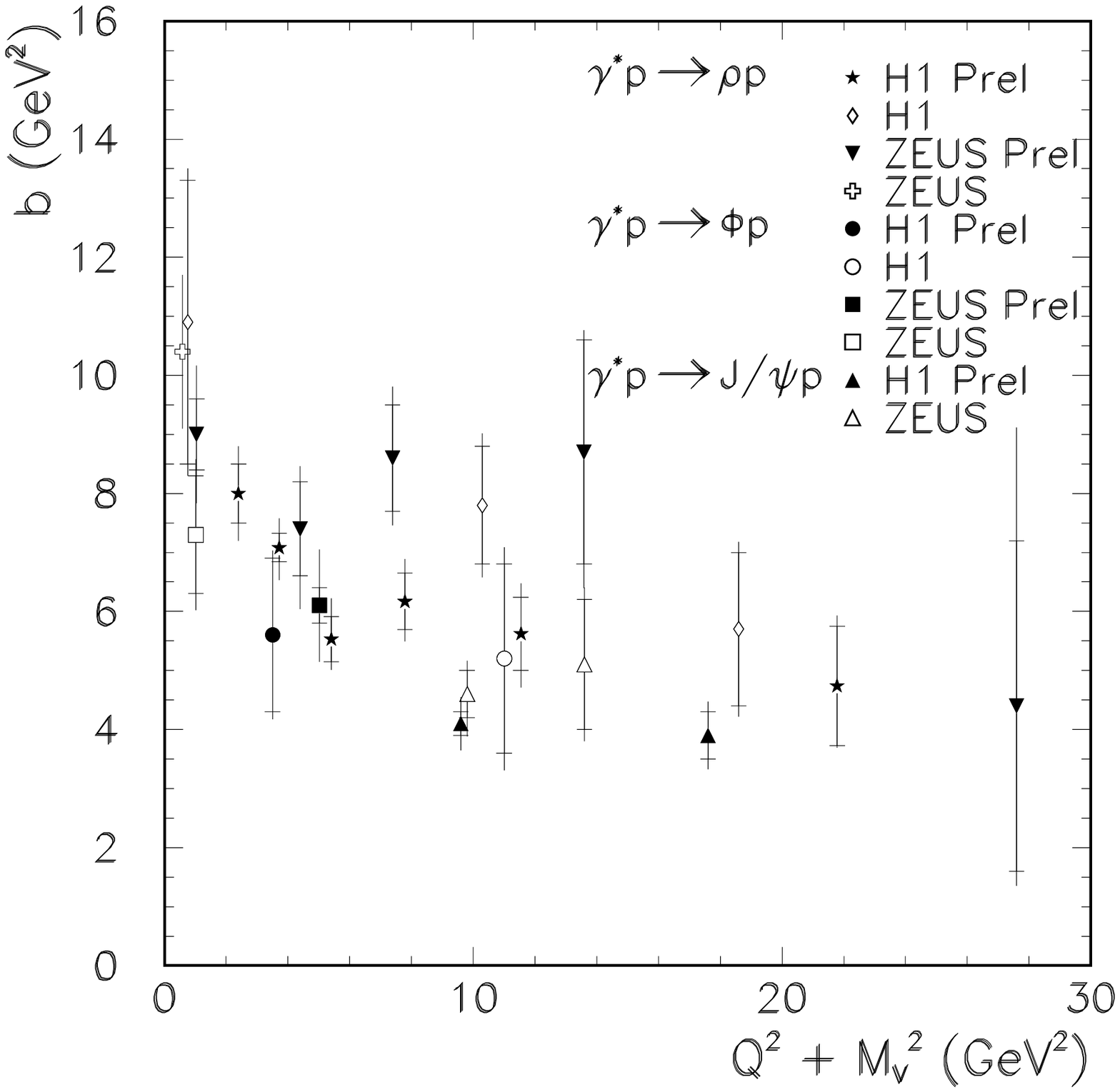,width=10cm}
\caption{
The fitted $b$ parameter of the four-momentum transfer $t$ distribution
$\exp{(-bt)}$ is shown for the production of different vector mesons in 
$ep$ collisions at HERA as a function of the scale which is here taken 
to be the sum of the photon virtuality $Q^2$ and the vector meson 
squared mass $M_V^2$.}
\label{fig:b-slope}
\end{figure}

In Fig.~\ref{fig:b-slope}, a compilation of the fitted $b$ parameters
is shown for the HERA data for $\rho$, $\varphi$, and $J/\psi$ meson
production as a function of the scale 
(new measurements: ~\cite{h1-rho,zeus-rho,zeus-phi,h1-psi}).
The scale has here again been chosen to be the sum of the photon virtuality
$Q^2$ and the vector meson squared mass $M_{V}^2$.
With increasing scale, the data tend to approach a constant value
of $b\sim 4$ GeV$^{-2}$ which corresponds to the size of the proton.
The size of the $q\bar{q}$ state is therefore small compared to that 
of the proton and probes the proton at small distances.

\begin{figure}[hht]
\setlength{\unitlength}{1cm}
\begin{picture}(5.0,13.5)
\put(0.0,-0.8)
{\epsfig{file=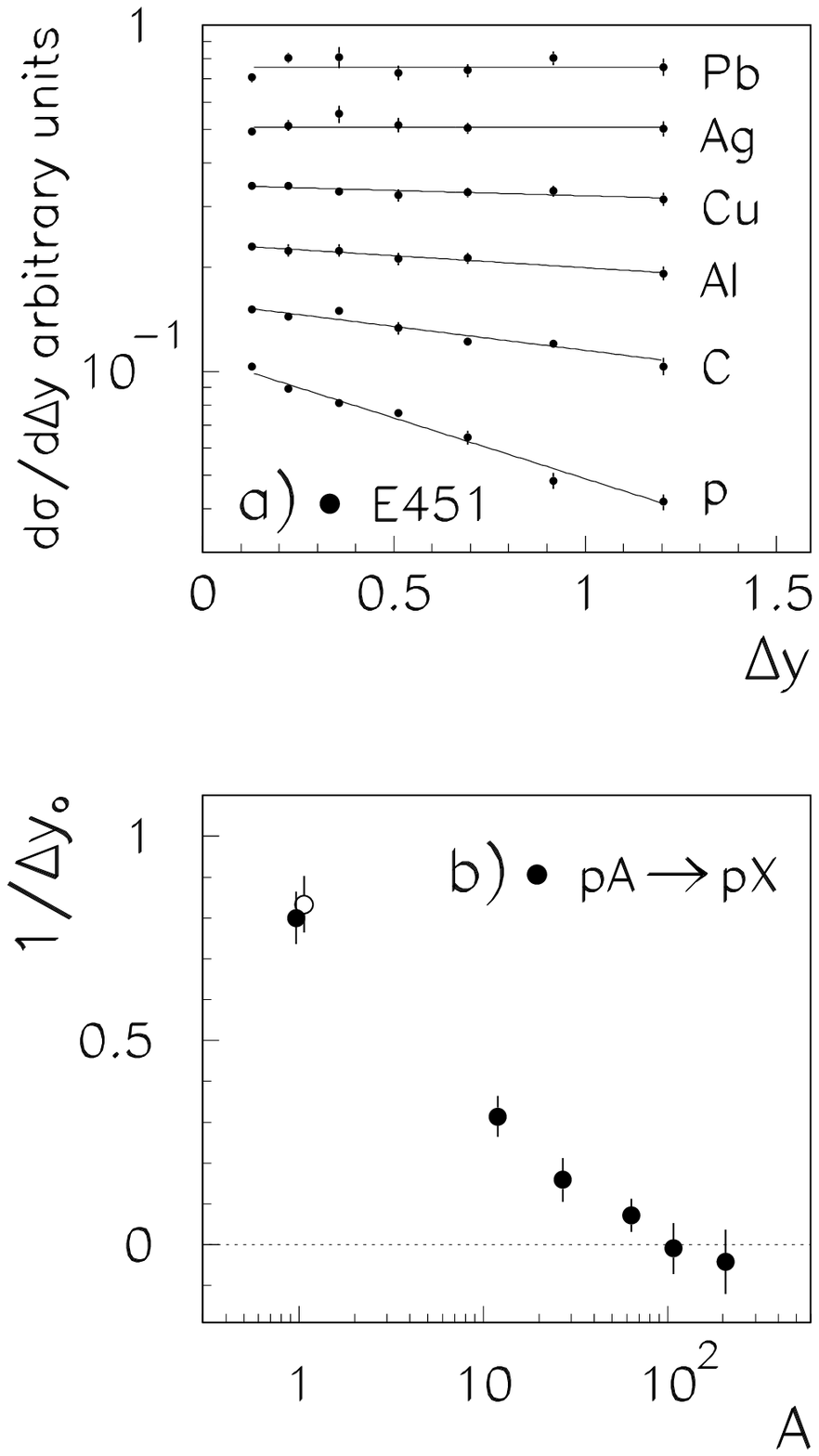,width=9cm}}
\put(8.0,-0.8)
{\epsfig{file=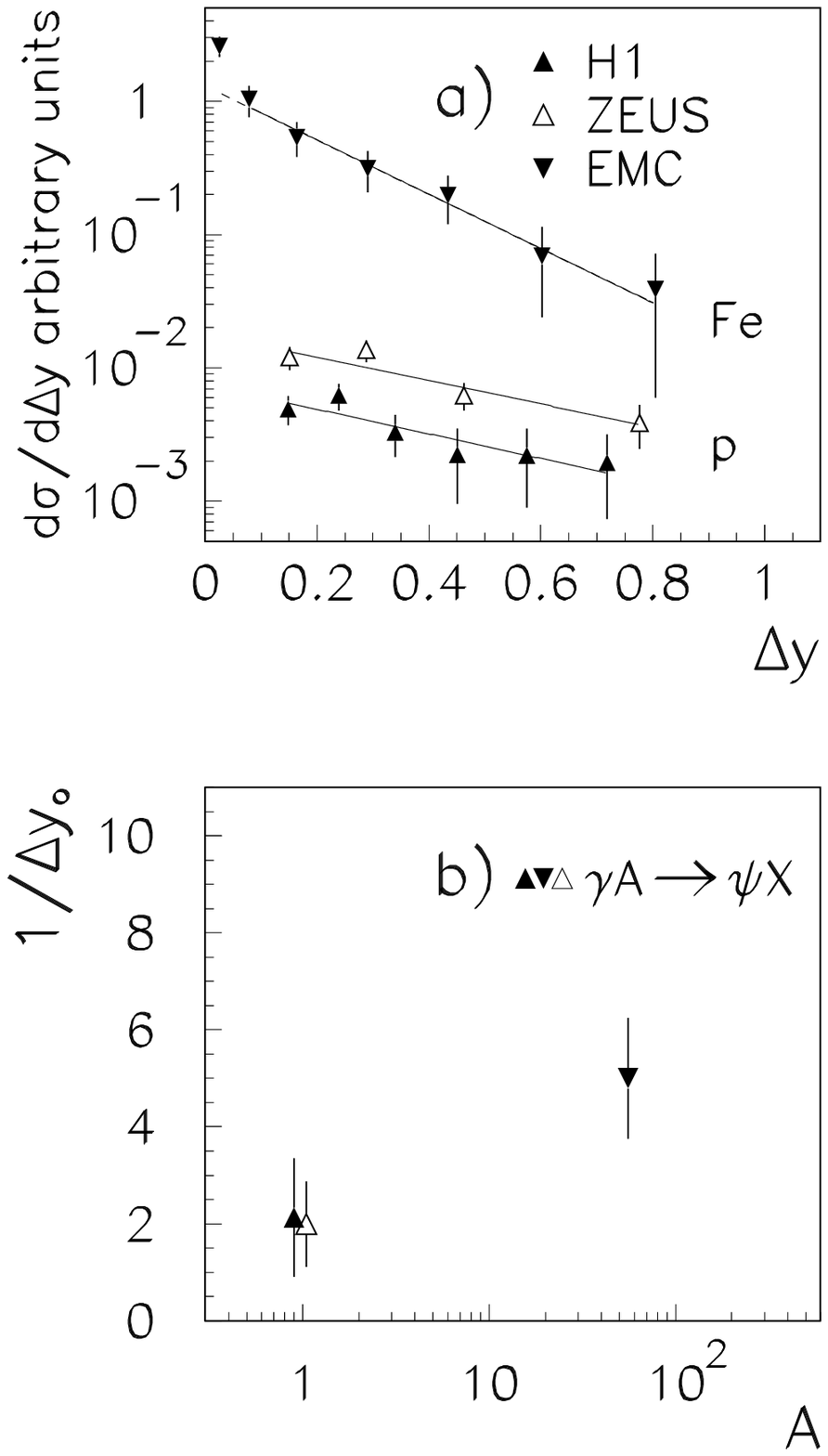,width=9cm}}
\end{picture}
\caption{
Left figures:
a) shape of the rapidity change of beam protons in proton--nucleus 
interactions.
The curves represent exponential fits to the data.
b) The slopes $1/\Delta y_\circ$ of the fits in a) are shown as a function of 
the target mass $A$.
Right figures:
a) shape of the rapidity difference between the 
initial-state photon and the $\psi$ meson
observed in lepton--proton (triangle symbols pointing up) and
lepton--iron interactions (triangle symbols pointing down).
The curves represent exponential fits to the data.
b) The slopes $1/\Delta y_\circ$ of the fits in c) are shown as a function 
of the target mass $A$.}
\label{fig:dy-proton} 
\end{figure}

The $b$ parameter measured for the photoproduction of 
$J/\psi$ mesons indicates the small 
size of the charm--anti-charm object in the interaction with the proton.
Further information on this $c\bar{c}$ configuration results from
nuclear dependencies of non-diffractive $J/\psi$ meson production in 
comparison to that of protons ~\cite{rapidity}:

In Fig.~\ref{fig:dy-proton}, the shapes of proton--nucleus cross sections 
$pA\rightarrow pX$ ~\cite{barton} are shown as a function of the rapidity change 
$\Delta y$.
Here $\Delta y$ denotes the rapidity difference between the beam proton
and the most energetic tagged proton.
These distributions can be described by an exponential form 
$\exp{(-\Delta y/\Delta y_\circ)}$.
The fitted slopes decrease as the nuclear mass increases, i.e.,
the protons are on average more decelerated with a heavier target.

It is interesting to compare the deceleration process of the protons 
with that of 
$J/\psi$ mesons resulting from non-diffractive photoproduction off nuclei
$\gamma A\rightarrow J/\psi X$.
Here the rapidity difference between
the photon and the $J/\psi$ is used as a measure of the 
deceleration process (Fig.~\ref{fig:dy-proton},
data of the EMC ~\cite{emc}, H1 ~\cite{h1-psi-old}, 
and ZEUS ~\cite{zeus-psi-old} experiments).

Also these distributions in the rapidity difference 
can be described by an exponential form.
In contrast to the proton data, the slope of the $J/\psi$ production 
does not decrease with increasing mass of the nucleus which implies that
the iron target does not decelerate the $c\bar{c}$ object better than 
the proton target.
The slight increase in the slope with $A$ can be explained,
according to Monte Carlo generator studies, by the different 
center-of-mass energies of the EMC and the HERA experiments.

The absence of a nuclear deceleration effect for the $J/\psi$ mesons may
be interpreted as resulting from nuclear transparency.
For a discussion of nuclear transparency effects refer, e.g., 
to ~\cite{frankfurt2}.
In this interpretation,
the colour charges of the small quark--anti-quark configuration are
sufficiently screened to penetrate a nucleus without further interactions.

\subsection{Summary 2: Colour Singlet Exchange}

\noindent
The HERA and Tevatron experiments have measured different observables 
that can be related to the parton distributions of diffractive exchange.
Using structure function measurements in $ep$ collisions, 
di-jet production in $ep$ and $\bar{p}p$ scattering, and
W-Boson production in $\bar{p}p$ collisions,
they consistently find a large gluon component in this colour singlet state.

The overall rate of diffractive processes observed in $ep$ and 
$\bar{p}p$ collisions, however, 
is found to be different and challenges explanation.

Measurements of elastic vector meson production involving a hard 
scale provide an alternative approach to understanding colour
singlet exchange.
Comparisons of the data with QCD calculations that rely on two-gluon
exchange give new information on the gluon distribution of the 
proton and on the vector meson states.

At sufficiently large scales, the spatial extension of the 
quark--anti-quark states appears to be small.
Photoproduction of $J/\psi$ mesons 
indicates that the $c\bar{c}$ state penetrates a nuclear environment 
essentially undisturbed.

Overall, diffractive physics is a very active field of research and
is developing away from a soft interaction language to the
understanding of a fundamental process of strong interactions within the 
framework of QCD.

\section*{Acknowledgements}
I wish to thank A. Astbury for providing a very positive conference
atmosphere for the exchange of the new scientific results.
I wish to thank for kind help in preparing the talk 
H.~Abramowicz, 
M.~Albrow,
V.~Andreev, 
K.~Borras, 
A.~Brandt, 
J.~Dainton, 
T.~Doyle,
K.~Freudenreich, 
C.~Glasman, 
B.~Heinemann, 
M.~Kienzle, 
P.~Newman, 
R.~Nisius,
G.~Snow, and
S.~S\"oldner-Rembold.
For careful reading of the manuscript and comments I wish to thank
H.~Abramowicz, 
M.~Albrow,
J.~Dainton, 
M.~Kienzle, 
P.~Newman,
G.~Snow, and
S.~S\"oldner-Rembold. \\
I am grateful to the Deutsche Forschungsgemeinschaft 
for the Heisenberg Fellowship.


\end{document}